\documentclass[08pt, a4paper, oneside]{article}
\usepackage[latin1]{inputenc}
\usepackage[english]{babel}
\usepackage{bm}
\usepackage{amsmath}
\usepackage{amsthm}
\usepackage{wasysym}
\usepackage{booktabs}
\usepackage{graphicx}
\usepackage{amstext}
\usepackage{color}
\usepackage{sidecap}
\usepackage{amsbsy,amssymb}
\numberwithin{equation}{section}

\newcommand{\CC}{\ensuremath{\mathbf C}}

\newcommand{\MM}{\ensuremath{\mathbf M}}

\renewcommand{\SS}{\ensuremath{\mathbf S}}

\begin{document}

\title{Modeling and analysis of water-hammer in coaxial pipes}

\author{Pierluigi Cesana\footnote{corresponding author, now at the Mathematical Institute, 
Woodstock Road, Oxford OX2 6GG, England; conducted theoretical analysis of the model},  Neal Bitter\footnote{prepared experimental set-up and performed experimental measurements}\\
\small California Institute of Technology, Pasadena, CA 91125, USA}

\maketitle

\begin{abstract}

The fluid-structure interaction is studied for a system composed of two coaxial pipes in an annular geometry, for both 
homogeneous isotropic metal pipes 
and 
fiber-reinforced (anisotropic) pipes. 
%
%The pipes are filled with water and a projectile causes an impact resulting in multiple waves %which travel at different speeds and amplitudes.
%
Multiple waves, traveling at different speeds and amplitudes, result when a projectile impacts on the water filling the annular space between the pipes.
In the case of carbon fiber-reinforced plastic thin pipes we compute the wavespeeds, the fluid pressure and mechanical strains as functions of the fiber winding angle. This generalizes the single-pipe analysis of J. H. You, and K. Inaba, \textit{Fluid-structure interaction in water-filled pipes of anisotropic composite materials}, J. Fl. Str. 36 (2013). 
Comparison with a set of experimental measurements seems to validate our models and predictions.

\end{abstract}


%\begin{center}
%\textbf{Preprint, submitted to : Journal of Fluids and Structures, December 2013 }
%\end{center}
%\tableofcontents

%\newpage

\section{Introduction}

This article is part of a series of papers \cite{HoYou13}, \cite{Bitter01}, \cite{Bitter02}, \cite{Inaba01}, \cite{Perotti01} devoted to the investigation of water-hammer problems in fluid-filled pipes, both from the experimental and theoretical perspective.
Water-hammer experiments are a prototype model for many situations in industrial and military applications (e.g., trans-ocean pipelines and communication networks) where we have fluid-structure interaction and a consequent propagation of shock-waves.
After the pioneering work of Korteweg$^{\cite{Korteweg}}$ (1878) and Joukowsky$^{\cite{Joukowsky}}$ (1900), who modeled water-hammer waves by neglecting inertia and bending stiffness of the pipe, a more comprehensive investigation, developed by Skalak \cite{Skalak} in the Fifties, considered inertial effects both in the pipe and the fluid, including longitudinal and bending stresses of the pipe. Skalak combined the Shell Theory for the tube deformation and an acoustic model of the fluid motion. He shows there is a coexistence of two waves traveling at different speeds: the precursor wave (of small amplitude and of speed close the sound speed of the pipe wall) and the primary wave (of larger amplitude and lower speed).
Additionally, a simplified four-equation one-dimensional model is derived based on the assumption that pressure and axial velocity of the fluid are constant across cross-sections \cite{Skalak}. Later studies of Tijsseling \cite{Tijsseling3}-\cite{Tijsseling} have regarded modeling of isotropic thin pipes including an analysis of the effect of thickness on isotropic pipes based on the four-equation model \cite{Tijsseling}.
While all these papers consider the case of elastically isotropic pipes, the investigation of anisotropy in water-filled pipes of composite materials was first obtained in \cite{HoYou13} where stress wave propagation is investigated for a system composed of 
water-filled thin pipe with symmetric winding angles $\pm\theta$. In the same geometry, a platform of numerical computations, based on the finite element method, was developed in \cite{Perotti01} to describe the fluid-structure interaction during shock-wave loading of a water-filled carbon-reinforced plastic (CFRP)
tube coupled with a solid-shell and a fluid solver.
More complex situations involve systems of pipes mounted coaxially where the annular regions between the pipes can be filled with fluid. 
In this scenario, B\"urmann has considered the modeling of non-stationary flow of compressible fluids in pipelines with several flow sections \cite{Burmann}. 
His approach
consists of reducing the system of partial differential equations governing the fluid-structure interaction in coaxial pipes into a 1-dimensional problem by the Method of Characteristics.
%Later works have appeared on the fluid-structure modeling in coaxial pipes based on %approaches in the time \cite{Cirovic} or frequency \cite{Levitsky} domain.
Later works have appeared on the
modeling of sound dispersion in a cylindrical viscous layer bounded by two elastic
thin-walled shells \cite{Levitsky} and of the wave propagation in coaxial pipes filled with either fluid or a viscoelastic solid \cite{Cirovic}.

%%%%%%%%%%%%%%%%%%%%%
%Later works have appeared on the wave propagation in coaxial pipes filled with fluid or a %viscoelastic solid as well \cite{Cirovic} and 

%\cite{Levitsky} 
%%%%%%%%%%%%%%%%%%%%%%%

%%%%%%%%%%%%%%%%%%%%%%%%%

Motivated by the recent experimental effort of J. Shepherd's group on the investigation of the water-hammer in annular geometries \cite{Beltman01}, \cite{Beltman02}, \cite{Bitter01}, \cite{Bitter02}, we extend the modeling work of \cite{Tijsseling} and \cite{HoYou13} to investigate the propagation of stress waves inside an annular geometry delimited by two water-filled coaxial pipes, in elastically isotropic and CFRP pipes.
A projectile impact causes propagation of a water pressure wave causing the deformation of the pipes. Positive extension in the radial direction of the outer pipe, accompanied by negative extension (contraction) in the radial direction of the internal pipe, causes an increase in the annular area thus activating the fluid-structure interaction mechanism.

The architecture of the paper is as follows. After reviewing the work of You and Inaba on the modeling of elastically anisotropic pipes, we present the six-equation one-dimensional model (Paragraph \ref{1311081717}) that rules the fluid-solid interaction in a two-pipe system. 
In Section \ref{1311081625} we compare our theoretical findings with experimental data obtained during a series of water-hammer experiments.
Finally, in the case of fiber reinforced pipes, the wave propagation and the computation of hoop and axial strain are described in full detail in Paragraph \ref{1312181514}.

\begin{table}[h!]
\centering
\caption{Notation. Here and in what follows subscript $i$ is either set to be equal to 1 (in the case in which we refer to the internal pipe) or 2 (external pipe).} \label{}
\begin{tabular}{|l p{4.81cm}|l p{5.37cm}|}
\hline $\overline{\CC_{i}}$  & stiffness matrix (x,y and z coordinates) & $u_{i,r},u_{i,z}$ & two-dimensional displacement components of pipe $i$ along $r$ and $z$ axis\\ 
\hline 
$\overline{\SS_{i}}$  & compliance matrix (x,y and z coordinates)
& $\dot{u}_{i,r},\dot{u}_{i,z} $ & two-dimensional velocity components of pipe $i$ along $r$ and $z$ axis \\
\hline $r,\varphi,z$  &  cylindrical coordinates  &  $\overline{\dot{u}}_{i,r},\overline{\dot{u}}_{i,z} $ & one-dimensional velocity components of pipe $i$ along $r$ and $z$ axis\\
\hline $t$  & time  & $\dot{u}_{i,z0} $ &magnitude of $\overline{\dot{u}}_{i,z}$  \\
\hline 
$V_f$ & volume fraction (fiber)
& $\theta$ & fiber winding angle \\
\hline $p(r,z,t)$ & two-dimensional fluid pressure   & $V$ & one-dimensional axial fluid velocity\\
\hline $P(z,t)$ & one-dimensional fluid pressure  & $V_0$ & magnitude of $V$\\
\hline $P_0$ & magnitude of $P$ & $v_{r},v_{z}$ & two-dimensional fluid velocity components  along $r$ and $z$ directions\\
\hline $P_{out}$ & pressure outside pipe 2  & $\varepsilon_{i},\gamma_i$ & normal and shear strain in pipe $i$\\
%\hline 
$P_{in}$ & pressure inside pipe 1 & &\\
\hline $R_i$  & inner radius of pipe $i$ & 
$c$ & wavespeed   
\\
\hline $e_{i}$  & thickness of pipe $i$ 
 &   $\sigma_{i},\tau_i$ & normal and shear stress \\
\hline  $\rho_{i,t}$ & density (pipe $i$)    &  $\sigma_{i,r},\sigma_{i,\varphi},\sigma_{i,z}$ & two-dimensional stress components along $r,\varphi$ and $z$ axis \\
\hline $\rho_w$ &density of fluid & $\overline{\sigma}_{i,z}$ & one-dimensional axial stress \\
\hline  
$K$ &bulk modulus of fluid 
& $\sigma_{i,z0}$ & magnitude of $\overline{\sigma}_{i,z}$ \\
\hline $E_{(1)}, E_{(3)}$  & effective Young's modulus along transverse and longitudinal directions in a single ply &   $E_{m}, E_f$  & Young's modulus for matrix and fiber \\
\hline 
$G_{31}$  & effective shear modulus in a single ply  &  
$G_{m}, G_f$  & shear modulus for matrix and fiber \\
\hline  $\nu_m,\nu_f$ & Poisson's ratio for matrix and fiber & $\rho_m, \rho_f $ & density of matrix and fiber   \\
\hline
\end{tabular}
%\caption{Notation. Here and in what follows subscript $i$ is either set to be equal to 1 (in the %case in which we refer to the internal pipe) or 2 (external pipe).} \label{}
\end{table}

\begin{figure}[h!]
\centering
\includegraphics[width=3.3cm]{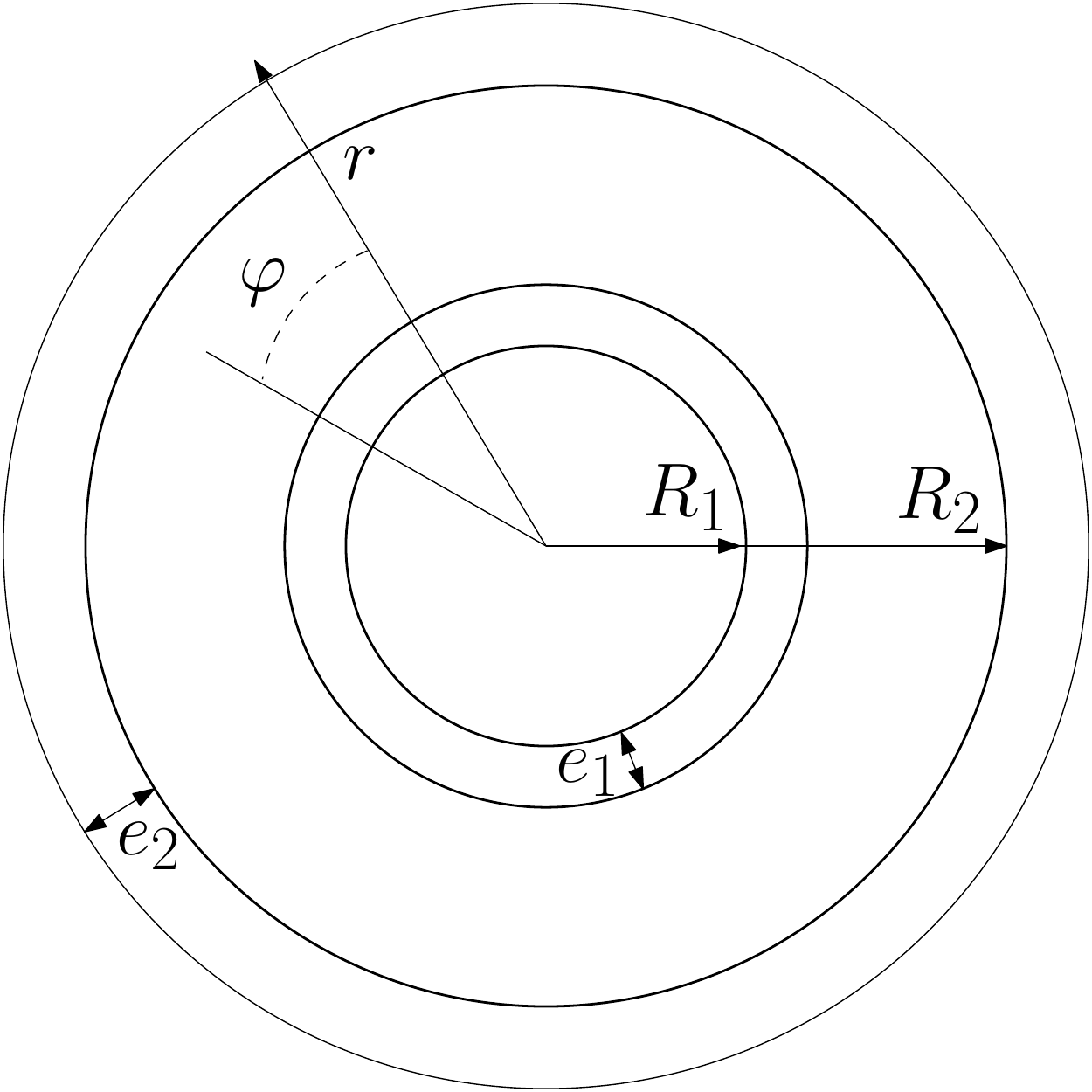}%
\qquad
\quad
%\raisebox{1pt}
\includegraphics[width=6.5cm]{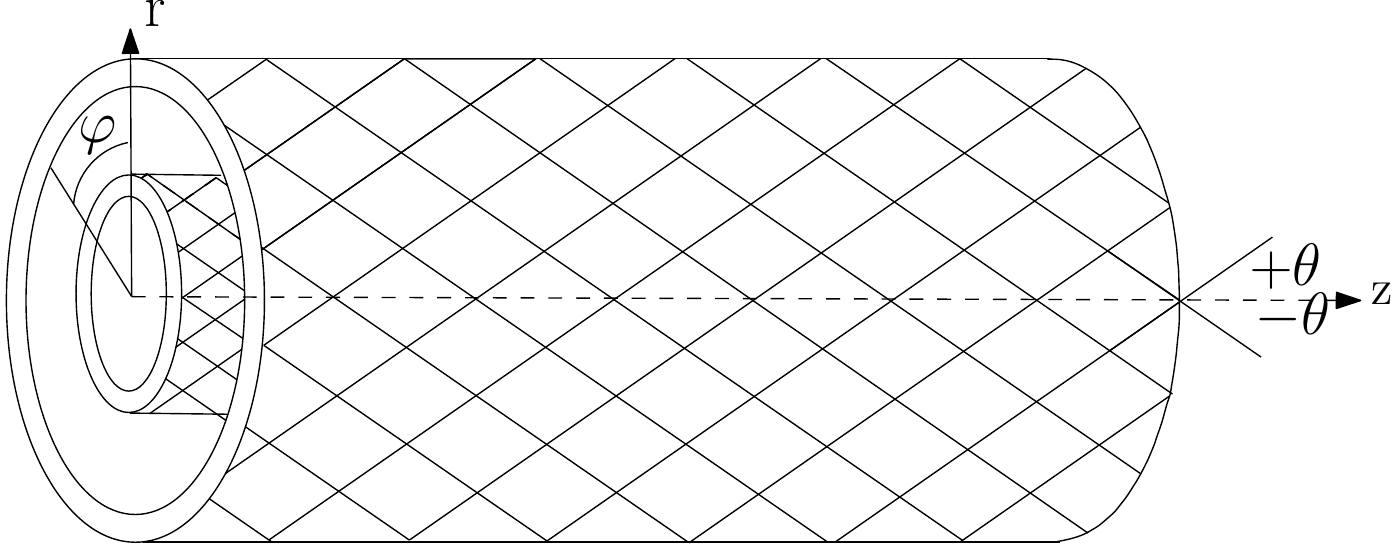}%
%\fbox{\includegraphics[width=5cm]{pipes5-eps-converted-to.pdf}
%\vspace{25mm}
\quad
\caption{Schematic representation of coaxial thin pipes. LEFT: cross section. RIGHT: lateral view. Notice that here we are referring to the case of CFRP pipes with winding angles $\pm\theta$. }\label{1309241710}
\end{figure}

\section{Thin pipes modeling}\label{13010041448}

\subsection{One-dimensional fluid-structure modeling}\label{1311081450}

According to the technique of Tijsseling \cite{Tijsseling}, one-dimensional governing equations for the liquid and the pipes can be obtained upon averaging out the standard balance laws in the radial direction. By adopting a cylindrical coordinates system, this approach is based upon the assumption that the behavior of water velocity and pressure depend only on the spatial variable $z$. In what follows we define one-dimensional cross-averaged quantities and obtain the corresponding field equations.

\subsubsection{Governing equations for the fluid}
The balance laws in the coordinate system $(r,z)$ for the fluid read \cite{Tijsseling3}
\begin{description}
\item[(2-d) Axial motion equation:\quad] 
$\displaystyle{\rho_w\frac{\partial v_z}{\partial t}+\frac{\partial p}{\partial z}=0},$
\item[(2-d) Radial motion equation:\quad] 
$\displaystyle{\rho_w\frac{\partial v_r}{\partial t}+\frac{\partial p}{\partial r}=0},$
\item[(2-d) Continuity equation:\quad] 
$\displaystyle{\frac{1}{K}\frac{\partial p}{\partial t}+\frac{\partial v_z}{\partial z}+\frac{1}{r}\frac{\partial(rv_r)}{\partial r}=0}.$
\end{description}
Here $v_z(r,z,t)$ and $v_r(r,z,t)$ are, respectively, the axial and radial velocity of the fluid and $p(r,z,t)$ is the pressure; $K$ is the bulk modulus of the fluid and $\rho_w$ is the density of the fluid.
We now introduce the cross-sectional averaged (one-dimensional) velocity and pressure, defined respectively as
\begin{eqnarray}
V(z,t):=\frac{1}{\pi\bigl(R_2^2-(R_1+e_1)^2\bigr)}\int_{R_1+e_1}^{R_2}2\pi r \,v_z(r,z,t)dr,\\
P(z,t):=\frac{1}{\pi\bigl(R_2^2-(R_1+e_1)^2\bigr)}\int_{R_1+e_1}^{R_2}2\pi r \,p(r,z,t)dr.
\end{eqnarray}
We are in a position to introduce the one-dimensional equations of balance for the fluid, which are
 \begin{description}
\item[(1-d) Axial motion equation] \hfill
\begin{eqnarray}\label{1305071550}
\rho_w\frac{\partial V}{\partial t}+\frac{\partial P}{\partial z}=0
\end{eqnarray}
\item[(1-d) Radial motion equation] \hfill
\begin{eqnarray}\label{1305071133}
\frac{1}{2}\rho_w  R_2\frac{\partial v_r}{dt}\Bigr|_{r=R_2}  +
\frac{R_2^2 p\Bigl|_{r=R_2}-(R_1+e_1)^2p\Bigl|_{r=R_1+e_1}}{\bigl( R_2^2-(R_1+e_1)^2\bigr)} -P=0
\end{eqnarray}
\item[(1-d) Continuity equation] \hfill
\begin{eqnarray}\label{1305071132}
 \frac{1}{K}\frac{\partial P}{\partial t}+\frac{\partial V}{\partial z}+
\frac{2}{(R_2^2-(R_1+e_1)^2)}\Bigl[R_2\, v_r\bigr|_{r=R_2}-(R_1+e_1)\, v_r\bigr|_{r=R_1+e_1}\Bigr]=0.
 \end{eqnarray}
\end{description}
We remark that Eq. (\ref{1305071133}) has been obtained by multiplying the two-dimensional radial motion equation by $2\pi r^2$, integrating in $r$ from $R_1+e_1$ to $R_2$ and dividing by $2\pi(R_2^2-(R_1+e_1)^2)$. Here $R_1$ is the internal radius and $e_1$ the thickness of the internal pipe while $R_2$ is the internal radius and $e_2$ the thickness of the external pipe (see Fig \ref{1309241710}-LEFT).
Moreover, in Eq. (\ref{1305071133}) it is assumed that  
\begin{eqnarray}
r\frac{\partial v_r}{dt}=R_2\frac{\partial v_r}{dt}\Bigr|_{r=R_2}=
(R_1+e_1)\frac{\partial v_r}{dt}\Bigr|_{r=R_1+e_1}.
\end{eqnarray}
This is consistent with the (2-d) Continuity equation under the hypothesis that $K$ is large and that the axial inflow $v_z$ is concentrated in the central axis in the limit $R_1\to 0$, $e_1/R_1\to 0$ \cite{Tijsseling}.

\subsubsection{Governing equations for the pipes}
Letting $i=1,2$, the equations of Axial motion and Radial motion in the pipes in the space $(r,z)$ are
\begin{description}
\item[(2-d) Axial motion equation:\quad] 
 $\displaystyle{\rho_{i,t}\frac{\partial \dot{u}_{i,z}}{\partial t}-\frac{\partial \sigma_{i,z}}{\partial z}=0},$
%\begin{eqnarray}\label{1302021614}
%\rho_{i,t}\frac{\partial \dot{u}_{i,z}}{\partial t}-\frac{\partial \sigma_{i,z}}{\partial z}=0,
%\end{eqnarray}
\end{description}
\begin{description}
\item[(2-d) Radial motion equation:\quad ] 
$\displaystyle{\rho_{i,t}\frac{\partial \dot{u}_{i,t}}{\partial t}=
\frac{1}{r}\frac{\partial (r\sigma_{i,r})}{\partial r}
- \frac{ \sigma_{i,\varphi}}{r}}.$
\end{description}
Here $\rho_{i,t}$ is the density of the pipe, $\sigma_{i,r}(r,z,t)$, $\sigma_{i,z}(r,z,t)$ and $\sigma_{i,\varphi}(r,z,t)$ are the radial, axial and hoop stress respectively and $\dot{u}_{i,r}$, $\dot{u}_{i,z}$ are the radial and axial velocity respectively.
By applying the cross-sectional average technique we obtain
\begin{description}
\item[(1-d) Axial motion equation] \hfill
\begin{eqnarray}\label{1305071551}
\rho_{i,t}\frac{\partial \overline{\dot{u}}_{i,z}}{\partial t}-\frac{\partial \overline{\sigma}_{i,z}}{\partial z}=0,
 \end{eqnarray}\end{description}
\begin{description}
\item[(1-d) Radial motion equation] \hfill
\begin{eqnarray}\label{1302021738}
\rho_{i,t}\frac{\partial \overline{\dot{u}}_{i,r}}{\partial t}=
 \frac{ (R_i+e_i)\sigma_{i,r}\Bigr|_{R_i+e_i} }{e_i(R_i+e_i/2)}
- \frac{ R_i\sigma_{i,r}\Bigr|_{R_i} }{e_i(R_i+e_i/2)}
-  \frac{1}{(R_i+e_i/2)}\overline{\overline{\sigma}}_{i,\varphi},
 \end{eqnarray}\end{description}
where
\begin{eqnarray}
\overline{\dot{u}}_{i,z}(z,t):=\frac{1}{\pi\bigl((R_i+e_i)^2-R_i^2\bigr)}\int_{R_i}^{R_i+e_i}2\pi r \,\dot{u}_{i,z}(r,z,t)dr,
 \end{eqnarray}
\begin{eqnarray}
\overline{\dot{u}}_{i,r}(z,t):=\frac{1}{\pi\bigl((R_i+e_i)^2-R_i^2\bigr)}\int_{R_i}^{R_i+e_i}2\pi r \,\dot{u}_{i,r}(r,z,t)dr,
 \end{eqnarray}
\begin{eqnarray}
\overline{\sigma}_{i,z}(z,t):=\frac{1}{\pi\bigl((R_i+e_i)^2-R_i^2\bigr)}\int_{R_i}^{R_i+e_i}2\pi r \,\sigma_{i,z}(r,z,t) dr,
 \end{eqnarray}
are respectively the one-dimensional axial velocity, radial velocity and axial stress and
\begin{eqnarray}
\overline{\overline{\sigma}}_{i,\varphi}:=\frac{1}{e_i}\int_{R_i}^{R_i+e_i}\sigma_{i,\varphi}(r,z,t)dr.
\end{eqnarray}

\subsubsection{Elastic properties of pipes}\label{1312181655}
 
By introducing the stiffness matrix $\overline{\CC}_i$ and the compliance matrix $\overline{\SS}_i:=\overline{\CC}_i^{-1}$, the stress-strain relation under the plane stress assumption reads, respectively, \cite[Eqs. (13, 14)]{HoYou13}
\begin{eqnarray}\label{13112331723}
%\begin{displaymath}
\left(
\begin{array}{ccc}
\sigma_{i,x} \\
\sigma_{i,z} \\
\tau_{i,zx}
\end{array} \right)=
\left(
\begin{array}{ccc}
\overline{C}_{i,11} &\overline{C}_{i,13} & 0 \\
\overline{C}_{i,13} & \overline{C}_{i,33} & 0 \\
0 & 0 & \overline{C}_{i,55}
\end{array} \right)
  \left(
\begin{array}{ccc}
\varepsilon_{i,x} \\
\varepsilon_{i,z} \\
\gamma_{i,zx}
\end{array} \right),\\
%\end{displaymath}
%or, equivalently,
%\begin{displaymath}
  \left(
\begin{array}{ccc}
\varepsilon_{i,x} \\
\varepsilon_{i,z}\\
\gamma_{i,zx}
\end{array} \right)
= \left(
\begin{array}{ccc}
\overline{S}_{i,11} &\overline{S}_{i,13} & 0 \\
\overline{S}_{i,13} & \overline{S}_{i,33} & 0 \\
0 & 0 & \overline{S}_{i,55}
\end{array} \right)
  \left(
\begin{array}{ccc}
\sigma_{i,x} \\
\sigma_{i,z} \\
\tau_{i,zx}
\end{array} \right).\nonumber
%\end{displaymath}
\end{eqnarray}
In the case of elastically homogenous and isotropic pipes, tensor $\overline{\CC}_i$ reads
\begin{eqnarray}\label{1305101405}
\left(
\begin{array}{ccc}
\overline{C}_{i,11} &\overline{C}_{i,13} & 0 \\
\overline{C}_{i,13} & \overline{C}_{i,33} & 0 \\
0 & 0 & \overline{C}_{i,55}
\end{array} \right)\equiv \left(
\begin{array}{ccc}
E_{i,t}/(1-\nu_{i,t}^2) & \nu_{i,t} E_{i,t}/(1-\nu_{i,t}^2) & 0 \\
\nu_{i,t} E_{i,t}/(1-\nu_{i,t}^2)  & E_{i,t}/(1-\nu_{i,t}^2)  & 0 \\
0 & 0 & G_{i,t}
\end{array} \right)
\end{eqnarray}
and, in turn
\begin{eqnarray}\label{1305101406}
\left(
\begin{array}{ccc}
\overline{S}_{i,11} &\overline{S}_{i,13} & 0 \\
\overline{S}_{i,13} & \overline{S}_{i,33} & 0 \\
0 & 0 & \overline{S}_{i,55}
\end{array} \right)= \left(
\begin{array}{ccc}
1/E_{i,t} & -\nu_{i,t}/E_{i,t} & 0 \\
-\nu_{i,t}/E_{i,t}   & 1/E_{i,t}  & 0 \\
0 & 0 & 1/G_{i,t}
\end{array} \right)
\end{eqnarray}
where $E_{i,t}$, $\nu_{i,t}$ and $G_{i,t}=E_{i,t}/(2+2\nu_{i,t})$ are the Young's modulus, Poisson's 
ratio and
shear modulus 
of the material from which pipe $i$ is made.

%\footnote{SINONIMO DI 3= longitudinal direction, SINONIMO DI 1= transverse direction}

\begin{SCfigure}
% \centering
\includegraphics[width=4cm]{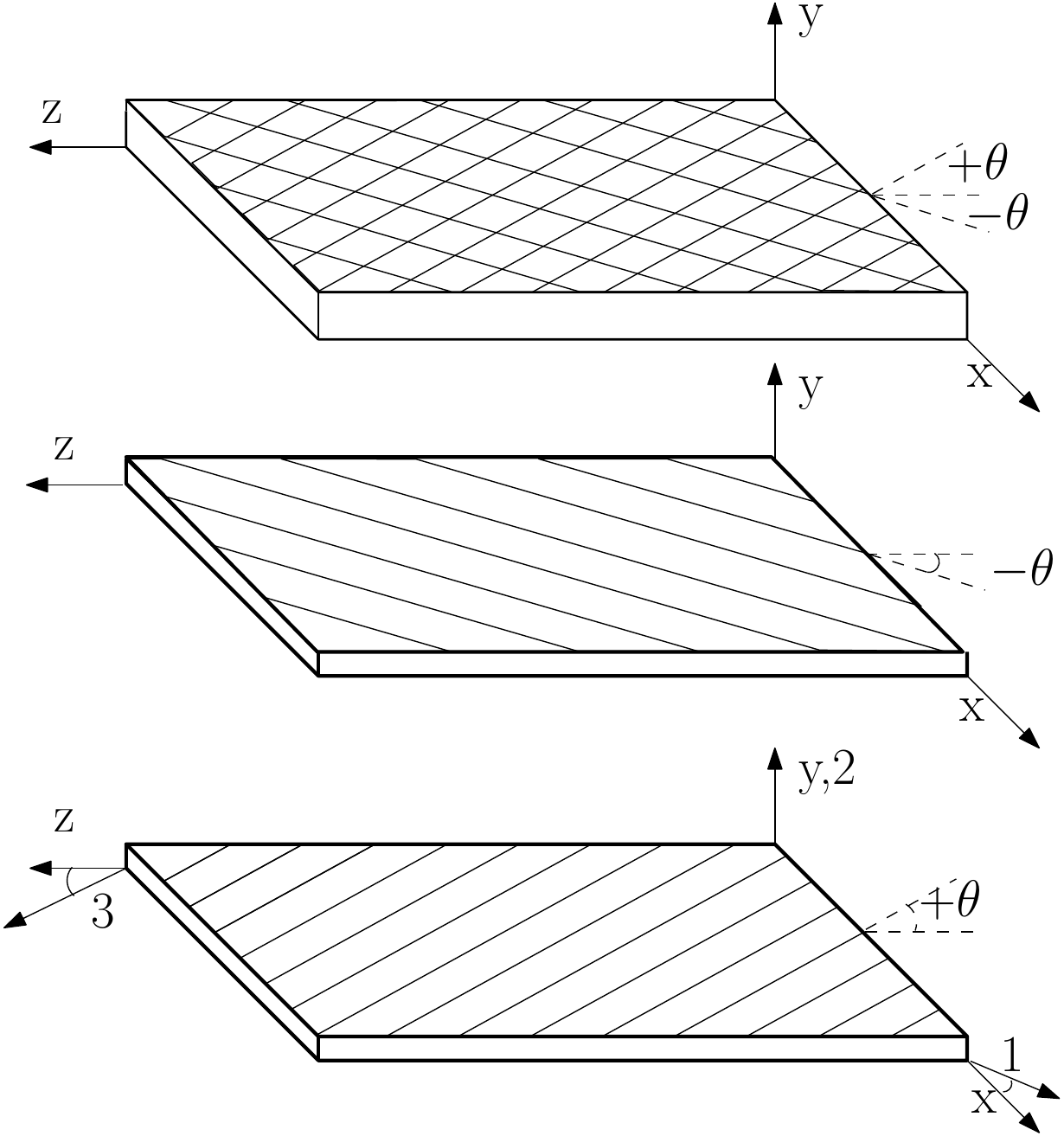}%
\caption{Schematic representation of CFRP lay-up structures as a combination of single uniaxial plies. Axes $1,2$ and $3$ are the principal axes of a single ply.}\label{1311101716}
\end{SCfigure}

For anisotropic composite (fiber-reinforced) pipes the stiffness elements $\overline{C}_{i,kl}$
are necessarily a function of the geometric and elastic properties of fibers and of matrix, including the fiber winding angle $\theta$. 
The difficulty in describing the elastic properties of fiber-reinforced plastic thin pipes has been studied in \cite{HoYou13} under the assumption that pipes are obtained by rolling up a woven layer with symmetric angles $\pm\theta$. Each of these layers can be considered as a lay-up structure of multiple plies of same thickness as shown in Figure \ref{1311101716}.
%
%
%
%We recall the stress-strain relation in the woven composite materials (Eq. (\ref{13112331723}))
%\begin{eqnarray}
%\left(
%\begin{array}{ccc}
%\sigma_{i,x} \\
%\sigma_{i,z} \\
%\tau_{i,zx}
%\end{array} \right)=
%\left(
%\begin{array}{ccc}
%\overline{C}_{i,11} &\overline{C}_{i,13} & 0 \\
%\overline{C}_{i,13} & \overline{C}_{i,33} & 0 \\
%0 & 0 & \overline{C}_{i,55}
%\end{array} \right)
 % \left(
%\begin{array}{ccc}
%\varepsilon_{i,x} \\
%\varepsilon_{i,z} \\
%\gamma_{i,zx}
%\end{array} \right).\nonumber
%\end{eqnarray}
%
%The stress-strain relationship for a single ply (of winding angle either $\theta$ or $-\theta$) is %derived under the plane stress assumption. 
%
%
To keep the notation simple, in what follows we drop the subscript $i$.
Elastic moduli are computed as a function of fiber volume fraction $V_f$ and fiber and matrix elastic coefficients \cite{KAW}
\begin{gather}
E_{(3)}=E_f V_f+E_m(1-V_f),\quad
\frac{1}{E_{(1)}}=\frac{ V_f}{E_f}+\frac{(1-V_f)}{E_m},\quad
\frac{1}{G_{31}}=\frac{V_f}{G_f}+\frac{(1-V_f)}{G_m},\label{1312091548}\\
\nu_{31}=\nu_f V_f+\nu_m(1-V_f),\qquad \rho_{i,t}=\rho_f V_f+\rho_m(1-V_f)\label{1312251840}
\end{gather}
where $E_m,G_m,\nu_m$ and $\rho_m$ are, respectively, the Young's modulus, shear modulus, Poisson's ratio and density of the matrix. Then $E_f,G_f,\nu_f$ and $\rho_f$ are, respectively, the fiber Young's modulus, shear modulus, Poisson's ratio and density of the fiber. The subscripts $3$ and $1$ indicate the longitudinal and transverse direction of a single ply (see Fig. \ref{1311101716}). The Poisson's ratio $\nu_{31}$ is defined
as the ratio of the contracted normal strain 
in the direction 1 to the normal strain in the direction 3, when a normal load is applied in the longitudinal direction.
%
%A woven layer with symmetric winding angles $\pm\theta$ consists of lay-up structure %consisting of a pair of $+\theta$ and $-\theta$ plies of equal thickness (see %\ref{1309241710}-RIGHT). 
%
The stiffness matrix for the composite is given by the volumetric average of the elastic stiffness matrices from each single $\pm \theta$ ply, denoted in what follows with $\overline{\CC}^{\pm\theta}$. 
Precisely, if all plies have the same thickness, the stiffness matrix for a woven layer of pairs of $\pm\theta$ plies is given by
\begin{eqnarray}\label{1312092133}
\overline{\CC}=\frac{1}{2}(\overline{\CC}^{+\theta}+\overline{\CC}^{-\theta}).
\end{eqnarray}
%
%(under the assumption that the plies have the same thickness)
%
%
The components of $\overline{\CC}^{\pm \theta}$ read \cite[Eq. (16)]{HoYou13}
\begin{eqnarray}
\begin{array}{llll}
\overline{C}_{11}^{+\theta}=\overline{C}_{11}^{-\theta} &= \bigl[ \cos^4(\theta) E_{(1)}/E_{\sharp} +  \sin^4(\theta)E_{(3)}/E_{\sharp}\bigr] +2\sin^2(\theta)\cos^2(\theta)\bigl[\nu_{31} E_{(1)}/E_{\sharp}+2G_{31}\bigr], \\[.2cm]
\overline{C}_{33}^{+\theta}=\overline{C}_{33}^{-\theta} &= [\cos^4(\theta)E_{(3)}/E_{\sharp} + \sin^4(\theta)E_{(1)}/E_{\sharp}] +2\sin^2(\theta)\cos^2(\theta)\bigl[\nu_{31} E_{(1)}/E_{\sharp}+2G_{31}\bigr], \\[.2cm]
\overline{C}_{13}^{+\theta}=\overline{C}_{13}^{-\theta} &= [ E_{(3)}/E_{\sharp} +E_{(1)}/E_{\sharp} -4 G_{31} -2\nu_{31} E_{(1)}/E_{\sharp} ]\sin^2(\theta)\cos^2(\theta) +\nu_{31} E_{(1)}/E_{\sharp},\\[.2cm]
\overline{C}_{55}^{+\theta}=\overline{C}_{55}^{-\theta} &=[
E_{(1)}/E_{\sharp}  -2 \nu_{31} E_{(1)}/E_{\sharp} + E_{(3)}/E_{\sharp}-4G_{31}] \cos^2(\theta) \sin^2(\theta) +G_{31} ,\\[.2cm]
\end{array}\nonumber
\end{eqnarray}
with $E_{\sharp}=1-\nu_{31}^2 E_{(1)}/E_{(3)}$. Finally, the compliance elements read $\overline{S}_{11}=\overline{C}_{33}/\overline{C}_{\sharp}$, $\overline{S}_{33}=\overline{C}_{11}/\overline{C}_{\sharp}$, $\overline{S}_{13}=-\overline{C}_{13}/\overline{C}_{\sharp}$ and $\overline{C}_{\sharp}=\overline{C}_{11}\overline{C}_{33}-\overline{C}_{13}^2$, where $\overline{C}_{kl}$ are the elements of the matrix $\overline{\CC}$ defined in (\ref{1312092133}).

\subsection{Six-equation model}\label{1311081717}

The axial and hoop strains in pipe $i$ can be written as \cite{HoYou13}
\begin{eqnarray}\label{1303281553}
\varepsilon_{i,z}=\overline{S}_{i,13}\sigma_{i,\varphi}+\overline{S}_{i,33}\sigma_{i,z},\quad
\varepsilon_{i,\varphi}=\overline{S}_{i,11}\sigma_{i,\varphi}+\overline{S}_{i,13}\sigma_{i,z},
\end{eqnarray}
respectively. By using the strain-displacements relations
\begin{eqnarray}\label{1311091742}
\varepsilon_{i,z}=\frac{\partial u_{i,z}}{\partial z},
\end{eqnarray} 
by differentiating in time and by taking the cross-sectional average, Eq. $(\ref{1303281553})$-LEFT becomes
\begin{eqnarray}\label{1302022228}
\frac{\partial \overline{\dot{u}}_{i,z}}{\partial z}=\overline{S}_{i,13}\frac{\partial\overline{\sigma}_{i,\varphi}}{\partial t}+\overline{S}_{i,33}\frac{\partial \overline{\sigma}_{i,z}}{\partial t}
\end{eqnarray}
where
\begin{eqnarray}
\overline{\sigma}_{i,\varphi}(z,t):=\frac{2\pi}{\pi\bigl((R_i+e_i)^2-R_i^2\bigr)}\int_{R_i}^{R_i+e_i}
r\,\sigma_{i,\varphi}(r,z,t)dr
\end{eqnarray}
is the one-dimensional (cross-averaged) hoop stress. 

The radial displacement equation is obtained by plugging another strain-displacement relation, which is,
\begin{eqnarray}\label{}
\varepsilon_{i,\varphi}=\frac{u_{i,r}}{r}
\end{eqnarray}
into Eq. (\ref{1303281553})-RIGHT yielding
\begin{eqnarray}\label{1305071058}
u_{i,r}=r \overline{S}_{i,11}\sigma_{i,\varphi}+r \overline{S}_{i,13}\sigma_{i,z}.
\end{eqnarray}
The equations of fluid and pipes are coupled by boundary conditions along the interfaces. 
Indeed, at each fluid-solid interface, we equate the radial velocity and radial stress of the fluid with those of the solid.
%\textsc{More precisely, we equate the fluid pressure with the radial pipes stress and we equate %the radial liquid velocity with the radial pipes velocity (no-slip condition) at both the interfaces %$r=R_1+e_1$ and $r=R_2$. Additionally, at the interface $r=R_2+e_2$ (respectively, $R_1$) we %impose the equality between the radial pipe stress and the external (respectively, internal) %pressure and between the radial pipe velocity and the velocity of the external (respectively, %internal) medium.}
\begin{eqnarray}\label{1309251040}
\begin{array}{llll}
     \sigma_{2,r}\bigr|_{r=R_2} &= -p\bigr|_{r=R_2} ,&\quad \dot{u}_{2,r}\bigr|_{r=R_2}&=v_r\bigr|_{r=R_2}, \\[.2cm]
     \sigma_{1,r}\bigr|_{r=R_1+e_1} &= -p\bigr|_{r=R_1+e_1} ,&\quad \dot{u}_{1,r}\bigr|_{r=R_1+e_1}&=v_r\bigr|_{r=R_1+e_1}, \\[.2cm]
      \sigma_{2,r}\bigr|_{r=R_2+e_2} &= -P^{out}=const. ,  &\quad \dot{u}_{2,r}\bigr|_{r=R_2+e_2}&=V_r^{out}=const. \,\,(=0 \,m/s),\\[.2cm]
\sigma_{1,r}\bigr|_{r=R_1} &=-P^{in} = const. , &\quad \dot{u}_{1,r}\bigr|_{r=R_1}&=V_r^{in}\,\,=const.\,\, (=0 \,m/s).\\[.2cm]
\end{array}
\end{eqnarray}
As in \cite{Tijsseling}, we assume that the external and internal pressures in each pipe induce a hoop stress which is constant in $\varphi$. Accordingly, we have \cite{HoYou13},  \cite{Timoshenko}
\begin{eqnarray}\label{1309251543}
\begin{array}{rcl}
\sigma_{1,\varphi}&=&\displaystyle{-\frac{1}{r^2} \frac{R_1^2(R_1+e_1)^2(P-P^{in} )}{2(R_1+e_1/2)e_1}+\frac{R_1^2P_{in}-(R_1+e_1)^2 P}{2(R_1+e_1/2)e_1}} \\[.4cm]
\sigma_{2,\varphi}&=&\displaystyle{-\frac{1}{r^2} \frac{R_2^2(R_2+e_2)^2(P_{out}-P)}{2(R_2+e_2/2)e_2}+\frac{R_2^2 P-(R_2+e_2)^2 P_{out}}{2(R_2+e_2/2)e_2}}. \\[.1cm]
\end{array}
\end{eqnarray} 
%and hence
%\begin{eqnarray}\label{1305071445}
%\begin{array}{rcl}
%\dot{\sigma}_{1,\varphi}&=&\displaystyle{-\frac{1}{r^2} \frac{R_1^2(R_1+e_1)^2(\dot{P})}%{2(R_1+e_1/2)e_1}+\frac{ -(R_1+e_1)^2 \dot{P}}%{2(R_1+e_1/2)e_1}} \\[.4cm]
%\dot{\sigma}_{2,\varphi}&=&\displaystyle{-\frac{1}{r^2} \frac{R_2^2(R_2+e_2)^2(-\dot{P})}%{2(R_2+e_2/2)e_2}+\frac{R_2^2\dot{P}}{2(R_2+e_2/2)e_2}.} %\\[.1cm]
%\end{array}
%\end{eqnarray} 
By plugging Eqs. (\ref{1309251040}-Lines 1 and 2, RIGHT) into Eq. (\ref{1305071132}) we obtain
\begin{eqnarray}\label{1305071407}
\frac{1}{K}\frac{\partial P}{\partial t}+\frac{\partial V}{\partial z}+
\frac{2}{(R_2^2-(R_1+e_1)^2)}\Bigl[R_2\, \dot{u}_{2,r}\Bigr|_{r=R_2}-(R_1+e_1)\, \dot{u}_{1,r}\Bigr|_{r=(R_1+e_1)}\Bigr]=0.
\end{eqnarray}
Now, assuming that radial inertial forces are ignored in both fluid and pipes and that the pipes cross-sections remain plane for axial stretches (thus implying the independency of $\sigma_{i,z}(z,t)$ on $r$, especially in thin pipes) we obtain a simplified model. Upon substitution of Eqs. (\ref{1305071058}) and (\ref{1309251543}) into Eq. (\ref{1305071407}) 
and upon substitution of Eq. (\ref{1309251543}) into (\ref{1302022228}) (by replacing $\sigma_{1,z}\bigr|_{r=R_1+e_1}=\overline{\sigma}_{1,z}$, $\sigma_{2,z}\bigr|_{r=R_2}=\overline{\sigma}_{2,z}$) we obtain new equations
%
%\begin{eqnarray}\label{1305071454}
%\frac{1}{K}\frac{\partial P}{\partial t}+\frac{\partial V}{\partial z}+
%\frac{2}{(R_2^2-(R_1+e_1)^2)}\Bigl[R_2^2\Bigl(\overline{S}_{2,11} \dot{\sigma}_{2,\varphi}\bigr|%_{r=R_2}+\overline{S}_{2,13}\dot{\sigma}_{2,z}\bigr|_{r=R_2}\Bigr)-\qquad\qquad\nonumber\\
%\qquad(R_1+e_1)^2\Bigl(\overline{S}_{1,11}\dot{\sigma}_{1,\varphi}\Bigr|%_{r=R_1+c1}+\overline{S}_{1,13}\dot{\sigma}_{1,z}\Bigr|_{r=R_1+c1}\Bigr)\Bigr]=0.
%\end{eqnarray}
%
\begin{eqnarray}\label{1305071553}
m_{21} \frac{\partial P}{\partial t}+\frac{\partial V}{\partial z}+m_{24}
\frac{\partial\overline{\sigma}_{2,z}}{\partial t}
-
m_{23}\frac{\partial\overline{\sigma}_{1,z}}{\partial t}=0,
\end{eqnarray}
and
\begin{eqnarray}\label{1305071555}
\frac{\partial \overline{\dot{u}}_{1,z}}{\partial z}=m_{51}\frac{\partial P}{\partial t}+\overline{S}_{1,33}\frac{\partial \overline{\sigma}_{1,z}}{\partial t},\qquad
\frac{\partial \overline{\dot{u}}_{2,z}}{\partial z}=m_{61}\frac{\partial P}{\partial t}+\overline{S}_{2,33}\frac{\partial \overline{\sigma}_{2,z}}{\partial t}
\end{eqnarray}
where
\begin{eqnarray}\label{1309302334}
\begin{array}{rcl}
  m_{21}&:=& \displaystyle{\frac{1}{K}}+\frac{\displaystyle{2\Bigl\{R_2^2\Bigl[\overline{S}_{2,11}\frac{(R_2+e_2)^2+R_2^2}{2(R_2+e_2/2)e_2}\Bigr]+  (R_1+e_1)^2\Bigl[\overline{S}_{1,11}\frac{R_1^2+(R_1+e_1)^2}{2(R_1+e_1/2)e_1}\Bigr]\Bigr\}}}{\displaystyle{(R_2^2-(R_1+e_1)^2)}}\\[.4cm]
&\approx& \displaystyle{\frac{1}{K}}+ \displaystyle{\frac{2}{R_2^2-(R_1+e_1)^2}\Bigl[\overline{S}_{2,11}\frac{R_2^3}{e_2}+   \overline{S}_{1,11}\frac{(R_1+e_1)^3}{e_1}\Bigr]} \\[.4cm]
  m_{23} &:=&\displaystyle{ \overline{S}_{1,13} \frac{2(R_1+e_1)^2}{R^2_2-(R_1+e_1)^2}} \\[.4cm]
    m_{24} &:=&\displaystyle{\overline{S}_{2,13} \frac{2R_2^2}{R^2_2-(R_1+e_1)^2}}\\[.4cm]  m_{51} &:=&\displaystyle{ \overline{S}_{1,13} H_1} \\[.4cm]
    m_{61} &:=&\displaystyle{\overline{S}_{2,13}H_2}\\[.4cm]
%    G&:=& \displaystyle{-\ln\Bigl(1+\frac{e_1}{R_1}\Bigr) \Bigl[\frac{(R_1+e_1)^2 %{2(R_1+e_1/2)e_1}\Bigr]\frac{2 R_1^2}{e_1(2R_1+e_1)}-\frac{ (R_1+e_1)^2  %{2(R_1+e_1/2)e_1}}\approx
%-\Bigl(\frac{R_1+e_1}{e_1}\Bigr)\\[.4cm]
%H&:=& \displaystyle{\ln\Bigl(1+\frac{e_2}{R_2}\Bigr) \Bigl[\frac{(R_2+e_2)^2 %{2(R_2+e_2/2)e_2}\Bigr]\frac{2 R_2^2}{e_2(2R_2+e_2)}+\frac{R_2^2  %{2(R_2+e_2/2)e_2}\approx\frac{R_2}{e_2}}\\[.4cm]
\end{array}
\end{eqnarray}
with
\begin{eqnarray}\label{1311101645}
\begin{array}{rcl}
    H_1&:=& \displaystyle{-\ln\Bigl(1+\frac{e_1}{R_1}\Bigr) \Bigl[\frac{(R_1+e_1)^2}{2(R_1+e_1/2)e_1}\Bigr]\frac{2 R_1^2}{e_1(2R_1+e_1)}-\frac{ (R_1+e_1)^2 }{2(R_1+e_1/2)e_1}}\approx
-\Bigl(\frac{R_1+e_1}{e_1}\Bigr)\\[.4cm]
H_2&:=& \displaystyle{\ln\Bigl(1+\frac{e_2}{R_2}\Bigr) \Bigl[\frac{(R_2+e_2)^2}{2(R_2+e_2/2)e_2}\Bigr]\frac{2 R_2^2}{e_2(2R_2+e_2)}+\frac{R_2^2 }{2(R_2+e_2/2)e_2}\approx\frac{R_2}{e_2}}.\\[.4cm]
\end{array}
\end{eqnarray}
The simplified expressions in Eqs. (\ref{1309302334}) and (\ref{1311101645}) are obtained under the assumption $(e_1/(R_1+e_1))\ll 1$, $e_2/R_2\ll 1$. 
Note that it is not possible for the terms in Eq. ((\ref{1309302334}) to become singular, since the denominator becomes zero only when the annulus of water has zero thickness.
%
%Throughout this paper we always assume that $R_2$ is sufficiently larger than $R_1+e_1$ so %that all the denominators in Eq. (\ref{1309302334}) are not singular.
%
Summarizing, the six-equations model with the six unknowns($P,V, \overline{\sigma}_{i,z}, \overline{\dot{u}}_{i,z}$) for the two-pipe system read
\begin{description}
  \item[Fluid (axial motion - continuity equation)] \hfill
\begin{eqnarray}\label{1305071618}
\rho_w\frac{\partial V_z}{\partial t}+\frac{\partial P}{\partial z}=0,\\
m_{21} \frac{\partial P}{\partial t}+\frac{\partial V}{\partial z}+m_{24}
\frac{\partial\overline{\sigma}_{2,z}}{\partial t}
-m_{23}\frac{\partial\overline{\sigma}_{1,z}}{\partial t}=0,
 \end{eqnarray}
\item[Pipes (axial motion - axial strain equation)] \hfill
\begin{eqnarray}
\rho_{i,t}\frac{\partial \overline{\dot{u}}_{i,z}}{\partial t}-\frac{\partial \overline{\sigma}_{i,z}}{\partial z}=0,\\
\frac{\partial \overline{\dot{u}}_{i,z}}{\partial z}-\overline{S}_{i,13}H_i\frac{\partial P}{\partial t}-\overline{S}_{i,33}\frac{\partial \overline{\sigma}_{i,z}}{\partial t}=0.
\label{1305071619}\end{eqnarray}
\end{description}
We seek solutions of (\ref{1305071618}-\ref{1305071619}) in the form of wave functions,
\begin{eqnarray}\label{1305071636}
P=P_0 f(z-ct), \quad V=V_0 f(z-ct), \quad
\overline{\sigma}_{i,z}=\sigma_{i,z0} f(z-ct), \quad\overline{\dot{u}}_{i,z}=\dot{u}_{i,z0} f(z-ct),
\end{eqnarray}
where $P_0,V_0,\sigma_{i,z0}$ and $\dot{u}_{i,z0}$ are magnitudes and $c$ is the wave speed. Substitution of (\ref{1305071636}) into (\ref{1305071618}-\ref{1305071619}) leads to six linear homogeneous equations which we write in a compact form
\begin{eqnarray}\label{1305091615}
\underbrace{\left[
\begin{array}{cccccc}
m_{11} & -c &0 &0 &0 &0   \\
-cm_{21} & 1 & cm_{23} &- cm_{24} &0 &0 \\
 0 &0 &-m_{33} &0 &  -c   & 0 \\
 0 & 0 &0 &-m_{44} &0 &  -c    \\
cm_{51} & 0 & cm_{53} & 0 & 1 &0 \\
cm_{61} & 0  & 0 & cm_{64} & 0 & 1
\end{array}\right]}_{\MM}
\left(
\begin{array}{c}
P_0  \\
V_0 \\
 \sigma_{1,z0}\\
 \sigma_{2,z0}\\
\dot{u}_{1,z0}\\
\dot{u}_{2,z0}\\
\end{array} \right)=
\left(
\begin{array}{c}
0  \\
0 \\
0\\
 0\\
0\\
0\\
\end{array} \right).
\end{eqnarray}
To keep a uniform notation we have defined $m_{11}=\rho_w^{-1}$, $m_{33}=\rho_{1,t}^{-1}$, $m_{44}=\rho_{2,t}^{-1}$, $m_{53}=\overline{S}_{1,33}$ and $m_{64}=\overline{S}_{2,33}$.
Existence of non-trivial solutions to (\ref{1305091615}) requires the determinant of $\MM$ to be zero, yielding, in turn, the following dispersion relation:
\begin{gather}
\bigl( m_{24} m_{53} m_{61} -  m_{23} m_{51} m_{64} - m_{21} m_{53} m_{64}\bigr)  c^6 +\nonumber
\\
 ( m_{21} m_{33} m_{64} +  m_{21} m_{44} m_{53} +
m_{23} m_{44} m_{51} -  m_{24} m_{33} m_{61} +  m_{11} m_{53} m_{64}  ) c^4 +\nonumber\\
(-m_{21} m_{33} m_{44} - m_{11} m_{33} m_{64} -  m_{11} m_{44} m_{53})\, c^2 +  m_{11} m_{33} m_{44}=0.\label{1305071655}
\end{gather}
Natural frequencies of the system $c_k$, with $k=1,...,6$, are the roots of (\ref{1305071655}). In general, Eq. (\ref{1305071655}) has to be solved by means of numerical methods. 
%
%However, if the pipes are composed of a matrix and fiber with the same volume fraction and %elastic properties (i.e., $\rho_{1,t}=\rho_{2,t}$, $\overline{S}_{1,13}=\overline{S}_{2,13}$ and  %$\overline{S}_{1,33}=\overline{S}_{2,33}$) 
%
However, if $\rho_{1,t}=\rho_{2,t}$, $\overline{S}_{1,13}=\overline{S}_{2,13}$ and  $\overline{S}_{1,33}=\overline{S}_{2,33}$
(e.g., if the pipes are composed of a matrix and fiber with the same volume fraction and elastic properties) 
we can find exact solutions of Eq. (\ref{1305071655}) analytically.
Indeed, if we define $p:=m_{33}/m_{53}=(\rho_{1,t}\overline{S}_{1,33})^{-1}$, $q:=m_{11}/m_{21}=(\rho_w m_{21})^{-1}$ and $\delta:=(m_{24}m_{61}-m_{23}m_{51})/(m_{53}m_{21})$, with $\delta\geq 0$ then Eq. (\ref{1305071655}) reads
\begin{eqnarray}
-(c^2-p)^2(c^2-q)+\delta c^4(c^2 -
 p)=0,
\end{eqnarray}
with roots
\begin{eqnarray}\label{1309302314}
\begin{array}{rcl}
      c_1&=&\sqrt{p},\label{1305072249} \\[.2cm]
      c_2&=&\displaystyle{\sqrt{\frac{(p+q)}{2(1-\delta)}+\sqrt{\frac{(p+q)^2}{4(1-\delta)^2} -\frac{4pq(1-\delta)}{4(1-\delta)^2}}}},\label{1304042252} \\[.5cm]
      c_3&=&\displaystyle{ \sqrt{\frac{(p+q)}{2(1-\delta)}- \sqrt{\frac{(p+q)^2}{4(1-\delta)^2} -\frac{4pq(1-\delta)}{4(1-\delta)^2}}}}:=c_w
\end{array}
\end{eqnarray}
$c_4=-c_1$, $c_5=-c_2$, $c_6=-c_3=-c_w$. Here $c_1,c_2$ and $c_w$ are positive (forward traveling) while $c_4,c_5$ and $c_6$ are negative (backward traveling) wave speeds. 
Since  $c_w$ is smaller than $c_1$ and $c_2$, we refer to it as the speed of the primary wave. Accordingly, we call
$c_1$ and $c_2$ the speeds of the precursor waves related to pipe 1 and 2, respectively.

\subsubsection{Reconstruction of the physical quantities}\label{1309301406}

We are now able to recover the mechanical strain in the hoop and axial directions as functions of $P,V,\overline{\sigma}_{i,z},\overline{\dot{u}}_{i,z} $. Thanks to Eq. $(\ref{1303281553})$-RIGHT we can write the one-dimensional hoop strain as follows
$$
\varepsilon_{i,hoop}=\overline{\varepsilon}_{i,\varphi}=\frac{1}{2\pi(R_i+e_i/2)e_i}\int_{R_i}^{R_i+e_i}2\pi r\, \varepsilon_{i,\varphi}dr=\overline{S}_{i,11}\overline{\sigma}_{i,\varphi}+\overline{S}_{i,13}\overline{\sigma}_{i,z},
$$
where $\overline{\sigma}_{i,\varphi}$ has been obtained in (\ref{1309251543}) (with $P_{in}= P_{out}=0$).
In turn, we have 
\begin{eqnarray}\label{1305101150}
\varepsilon_{i,hoop}=\overline{S}_{i,11}H_i P_0 f(z-ct)+\overline{S}_{i,13}\sigma_{i,z0}f(z-ct). 
\end{eqnarray}
The cross-sectional averaged axial strain can be obtained by Eqs. (\ref{1311091742}) and (\ref{1305071619})
\begin{eqnarray}
\varepsilon_{i,ax}=\overline{\varepsilon}_{i,z}=\frac{1}{2\pi(R_i+e_i/2)e_i}\int_{R_i}^{R_i+e_i}2\pi r\varepsilon_{i,z}dr=\frac{\partial \overline{u}_{i,z}}{\partial  z}
=-\frac{\dot{u}_{i,z0} }{c}f(z-ct)=\nonumber\\
\overline{S}_{i,13}H_i P_0 f(z-ct)+\overline{S}_{i,33}\sigma_{i,z0}f(z-ct).\label{1309251620}
\end{eqnarray}
Notice that here $ \overline{u}_{i,z}=(-\dot{u}_{i,z0} /c)F(z-ct)$ follows from integrating the last equation in (\ref{1305071636}) over time with $F'=f$. 
Then, from the system of equations (\ref{1305091615}) we can easily derive the following relations
\begin{eqnarray}\label{1309291450}
\dot{u}_{i,z0} =\frac{c \overline{S}_{i,13} H_iP_0}{\displaystyle{(-1+c^2\overline{S}_{i,33}\rho_{i,t})}},\qquad
\sigma_{i,z0} =-c\rho_{i,t}\frac{c \overline{S}_{i,13}H_i P_0}{ \displaystyle{(-1+c^2\overline{S}_{i,33}\rho_{i,t})}},\label{1303281900}
\end{eqnarray}
and, by taking $c=c_3=c_w$,
\begin{eqnarray}\label{1305091618}
P_0=c_w\rho_w V_0.
\end{eqnarray}
Thanks to Eq. (\ref{1305091618}) we can express averaged hoop and axial stress dependent on either the fluid velocity $V_0$ or, upon inversion of Eq. (\ref{1305091618}), on the fluid pressure $P_0$.
By plugging Eqs. (\ref{1303281900})-RIGHT and (\ref{1305091618}) into Eq. (\ref{1305101150}) and (\ref{1309251620}) with $c=c_w$ we obtain
\begin{gather}
\varepsilon_{i,hoop}=\Bigl\{\overline{S}_{i,11}H_i   +\overline{S}_{i,13}\Bigl[-c_w\rho_{i,t}\frac{c_w \overline{S}_{i,13}H_i}{(-1+c_w^2 \overline{S}_{i,33}\rho_{i,t})}\Bigr]\Bigr\}c_w\rho_w V_0 f(z-c_wt),\label{1311100014}\\
\varepsilon_{i,ax}=-\frac{\overline{S}_{i,13}H_i}{(-1+c_w^2 \overline{S}_{i,33}\rho_{i,t})} c_w\rho_w V_0 f(z-c_w t),\label{1309291512}\\
\sigma_{i,z0} =-\rho_{i,t}c_w^2\frac{\overline{S}_{i,13}H_i}{ \displaystyle{(-1+c_w^2\overline{S}_{i,33}\rho_{i,t})}}c_w\rho_w V_0.\nonumber
\end{gather}

\section{Water-hammer experiments}\label{1311081625}

Armed with the set of analytic expressions from Section \ref{13010041448}, we turn now to the simulation of experimental measurements for the water-hammer experiment for a set of pipes including homogeneous (isotropic) metal pipes and fiber-reinforced (anisotropic) pipes. 

\subsection{Experimental setup}
\begin{figure}[t]
\centering
\includegraphics[width=0.7\textwidth]{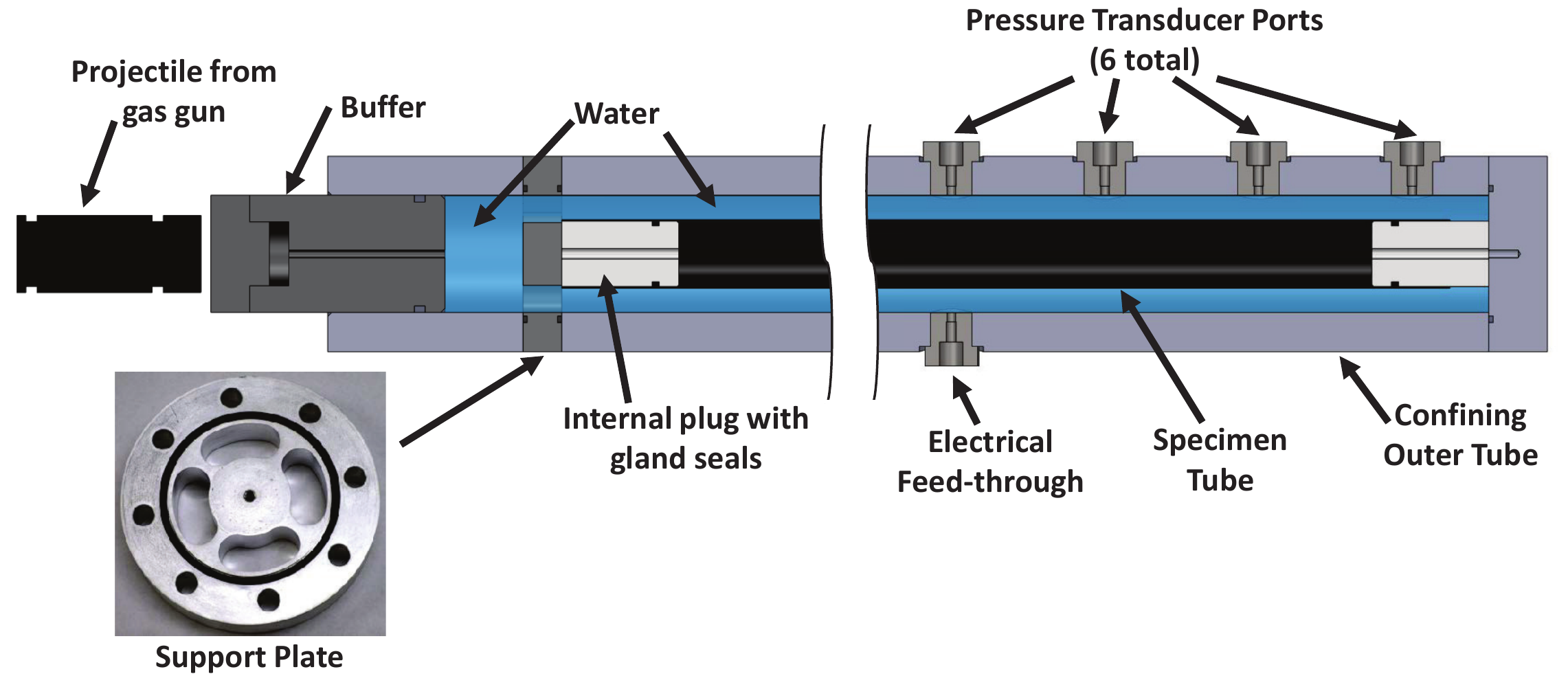}
\caption{Diagram of apparatus used to measure wave propagation speeds (rotated 90$^\circ$ counterclockwise). ``Electrical feedthrough'' provides a sealed connection between the leads of strain gauges (attached to specimen tube) and the data acquisition electronics external to the setup. The web version of this article contains the above plot figure in color.}
\label{fig:experiment}
\end{figure}

The propagation of waves in the annular space between two pipes is studied experimentally using the apparatus shown in Fig.~\ref{fig:experiment}. For all experiments, the outer pipe is a thick-walled cylindrical vessel made from 4140 high strength steel with an inner radius  $R_2=$ 38.1~mm, wall thickness $e_2=$ 25.4~mm, and length 0.97~m. Tubes of various sizes can be mounted concentrically inside of this vessel; these tubes are held in place at both ends by polycarbonate plugs and sealed with gland seals. The plug at the right end of Fig.~\ref{fig:experiment} is fixed to the base of the setup while the plug at the left end is fixed to a ``support plate'' (Fig.~\ref{fig:experiment} inset) which features four holes that allow pressure waves to pass freely while still providing support for the inner tube. 

The annular space between the two pipes is filled with distilled water while the sealed cavity inside of the inner pipe remains filled with air at ambient pressure. The water inside the annular cavity is also at atmospheric pressure at the beginning of the test. An aluminum ``buffer'' is then inserted into the top of the apparatus, and any air that remains in the system is vented through a hole in the buffer that is later sealed. The buffer is composed of a 125~mm long aluminum cylinder capped by a 25~mm thick steel striker plate which prevents damage of the buffer during projectile impact; this striker plate is bolted to the aluminum buffer using eight 1/4-20 machine screws.

\begin{figure}
\centering
\includegraphics[width=0.7\textwidth]{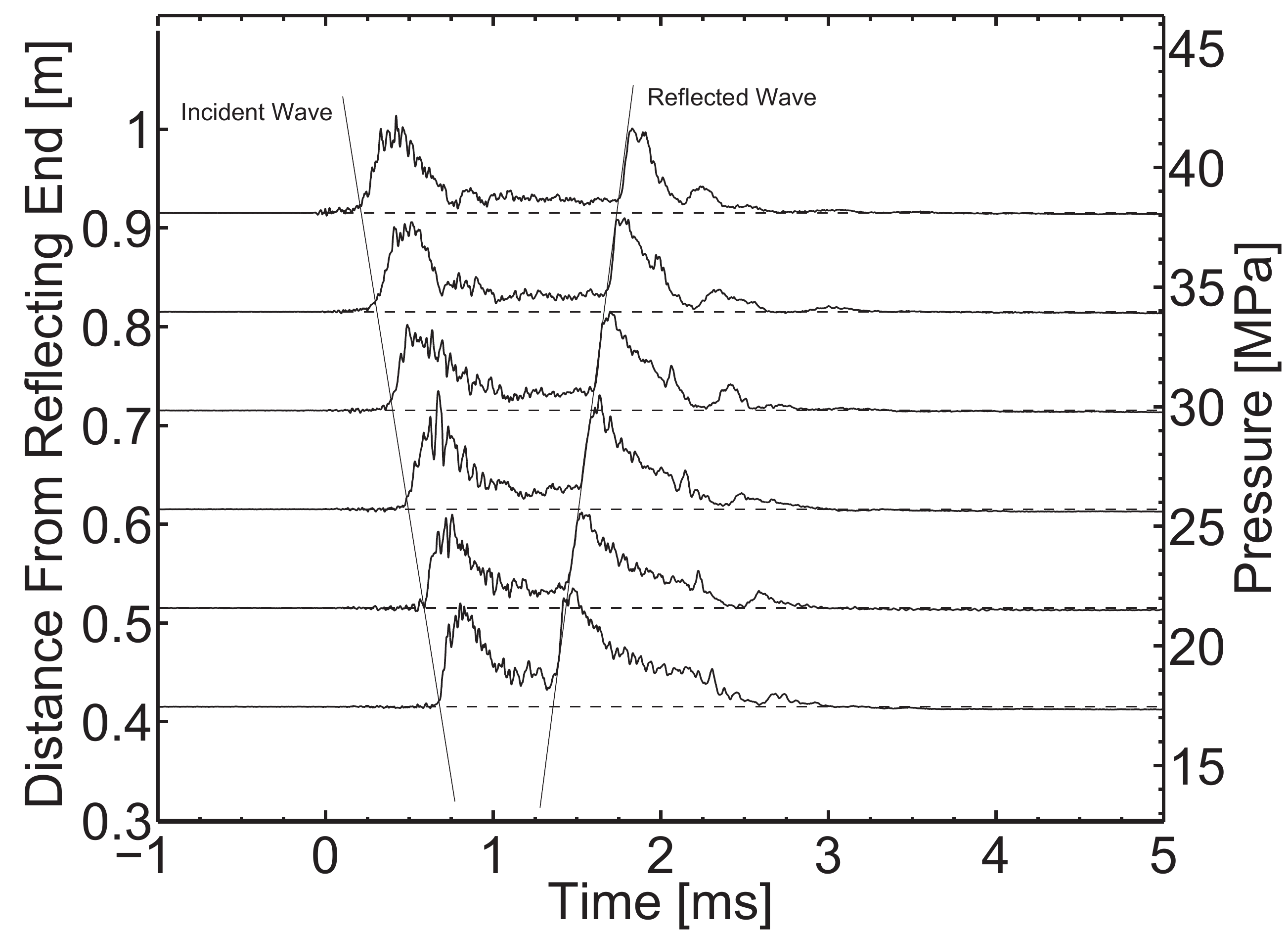}
\caption{Example of experimental pressure traces used to determine wave propagation speed. }
\label{fig:pressureTrace}
\end{figure}

Next a projectile is fired from a gas gun into the buffer, and the stress wave that develops in the buffer is transmitted into the water as a shock wave which travels along the annular space between the inner and outer pipes. Examples of pressure traces plotted on an $x-t$ diagram are shown in Fig.~\ref{fig:pressureTrace}; they demonstrate a sharp shock wave followed by an approximately exponential decay. Slight steepening of the wavefront is visible as the wave progresses. The high frequency fluctuations in the pressure signal are the result of axisymmetric vibrations of the inner pipe, as was confirmed by hoop strain measurements (not shown) at multiple locations around the circumference. 

The axial propagation speed of the pressure wave is measured using a row of six pressure transducers, PCB model 113A23, which have a response time less than 1~$\mu s$, resonant frequency above 500~kHz, and are sampled at 1~MHz. These transducers are mounted flush with the inner surface of the outer pipe. The response of the tube is recorded using bonded strain gauges oriented in the hoop direction, which are coated with a compliant sealant (Vishay PG, M-Coat D) to avoid electrical interference from the water. Signals are amplified using Vishay 2310B signal conditioners and digitized at 1~MHz.  

The speed of the pressure wave is measured by marking the time of arrival of the pressure wave at each transducer and fitting a line to the data as shown in Fig.~\ref{fig:pressureTrace}; the slope of this line is the wave speed. The time of arrival of the pressure wave at each transducer is determined as the first instant at which the pressure exceeds a chosen threshold. Because the shock wave steepens as it progresses, the wave speed depends slightly on chosen threshold level. For the results reported in this paper, wave speeds were calculated using both 50\% and 80\% of the maximum pressure of the incident wave as threshold values, and in every case the results differed by less than 5\%. The lower threshold value of 50\% was chosen because it is larger than the amplitude of the precursor wave, which ensures that the measured primary wave speed is not influenced by that of the precursor wave.

For each specimen tube, between 3 and 8 shots were conducted. After each shot, wave speeds were calculated using both 50\% and 80\% of the maximum pressure as the threshold described above, and the average of these two values was taken as the measured wave speed for the shot. Finally, the average wave speed over all 3-8 shots was calculated, and this number was taken to be the wave speed associated with the specimen. The uncertainty in the measured wave speed was calculated as twice the standard deviation of the measured wave speeds over all 3-8 shots; this uncertainty is typically less than 10\%. The difference in measured wave speed between shots depends mainly on the quality of the impact between the projectile and buffer. Slight non-normal impact is unavoidable and leads to slight differences in the measured velocity. 

Further details about the experimental setup are recorded in Refs. \cite{Bitter01,Bitter02}. It may be noted that the experiment was designed for the study of plastic deformation and buckling of tubes loaded by dynamic external pressure, which is the reason for the rather high pressure of the shock wave (on the order of 1-10 MPa). For all results reported in this paper, the pressure and total impulse of the pressure load were low enough to prevent plastic deformation of both the inner and outer pipes, as was verified by hoop strain measurements. 
Although the hoop strain measurements indicated slight non-axisymmetric deformation of the inner pipe (elastic buckling) in most shots, these effects do not appear to significantly influence the propagation of pressure waves so long as the buckles remain elastic (see Ref. \cite{Bitter01}).

%
%
%\begin{table}[h]
%\centering
%\begin{tabular}{|c|c|c|c|c|}\hline
%$R_2=38.1$ mm  &    $e_2=25.4$ mm & $E_{2,t}=198$ GPA & $\rho_{2,t}=7800$ kg/m$^3$ &  %$\nu_{2,t}=0.3$ \\
%\hline
%\end{tabular}
%\caption{Parameters of the large 4140 steel tube (referred as pipe 2 in Section %\ref{1311081625}).}\label{1311101234}
%\end{table}
%
%
\begin{table}[h!]
\centering
\caption{Parameters of the large 4140 steel tube (referred as pipe 2 in Section \ref{1311081625}) and water.}\label{1311101234}
\begin{tabular}{llcccccc}
\hline
%&\multicolumn{2}{l}{Measurements} 
%&\multicolumn{5}{r}{Computed quantities} 
%\\
%\cline{2-4}
%\cline{6-8}
4140 Steel  (pipe 2)
%  & $P_0$  & $\varepsilon_{1,hoop}$   & $V_0^{\dagger}$ & &$\varepsilon_{1,hoop}^{\sharp}$ & % $V_0^{\sharp}$ & $\varepsilon_{1,hoop}^{\ddagger}$\\
 % & [MPa]  &[mstr]  & [m/s] & & [mstr] & [m/s] &[mstr]
\\
\hline
Young's modulus      & $E_{2,t}$  & 198 GPa    &  &  & Inner radius   & $R_2$  & 38.1 mm\\
Density      & $\rho_{2,t}$   & 7800 kg/m$^3$    & &  & Thickness  & $e_2$ & 25.4 mm\\
Poisson's ratio       & $\nu_{2,t}$   & 0.3   &    & &   & &  \\
\hline
Water\\
\hline
Bulk modulus    & $K$     & 2.14 GPa     & &  &   &  &  \\
Density    & $\rho_w$    & 999 kg/m$^3$   & & &  &  &  \\
\hline
\end{tabular}
%\caption{Parameters of the large 4140 steel tube (referred as pipe 2 in Section %\ref{1311081625}) and water.}\label{1311101234}
\end{table}

\begin{table}[h!]
\centering
\caption{Parameters of stainless steel and aluminum.}\label{1312281935}
\begin{tabular}{llcclcc}
\hline
Aluminum (pipe 1) &  & &   & Stainless steel (pipe 1) &  & 
\\
\hline
Young's modulus  & $E_{1,t}$  & 68.9 GPa    &    & Young's modulus      & $E_{1,t}$  & 198 GPa\\
Density  & $\rho_{1,t}$   &  2700 kg/m$^3$     &  & Density  & $\rho_{1,t}$ &  8040 kg/m$^3$ \\
Poisson's ratio       & $\nu_{1,t}$   & 0.33   &     & Poisson's ratio  & $\nu_{1,t}$ & 0.29 \\
\hline
\end{tabular}
%\caption{Parameters of stainless steel and aluminum.}\label{1312281935}
\end{table}

\subsection{Results and discussion}\label{1312181656}

We now address our experimental measurements of the speed of the primary wave, the fluid pressure and the hoop strain in the internal pipe against the prediction based on our modeling work.
The geometrical and physical properties of the external pipe are reported in Table \ref{1311101234}.
The smaller interior tubes are mounted concentrically and are made from either aluminum, stainless steel (with properties listed in Tables \ref{1312281935} and  \ref{1311101332}) or a carbon fiber-epoxy resin matrix composite.
The elastic coefficients matrices $\overline{\CC}_i$ and $\overline{\SS}_i$ for both the internal ($i=1$) and external ($i=2$) pipe are computed as in Paragraph \ref{1312181655} with the only exception being the CFRP pipe for which a clarification of the method is given in the Remark below.
Notice that since the physical properties of the external and internal pipes are different, and in particular $\overline{\CC}_1\neq\overline{\CC}_2$, computation of the primary wave speed $c_w$ needs to be accomplished by solving Eq. (\ref{1305071655}) numerically. Results displayed in Table \ref{1311101332} show a good match
between the experimental data and the predicted values. However, there are some slight differences; in particular, the measured wave speeds are consistently greater than those predicted by the model. One possible explanation is that the steepening wavefront of the pressure wave biases the experimental measurements toward higher wave speeds.

%\textbf{This is particularly remarkable if we consider that the ratio $R_2/e_2$ is close to the %identity, \textbf{thus leaving the assumptions of thin pipes unfulfilled. }}

Comparisons of experimental measurements and predictions for
hoop and axial strain in the internal pipe, given in Table~\ref{1310091629}, show that the calculated data match with experimental results reasonably well. In this table the measured data points are the peak values of the pressure and hoop strain, which were determined after applying a 50~kHz low-pass filter to the recorded signals. 
%The computed values of these physical variables are based on the primary wave speed, $c_w$, %since slower waves generate larger magnitudes in both axial and hoop strain. 
The computed values of these physical variables are based on $c_w$ since the primary wave generates larger magnitudes in both axial and hoop strain than the precursor waves.
Differences between the computed and measured values are the result of modeling idealizations that are not completely satisfied in the experiment. For instance, radial inertia of the fluid and pipes,  strain rate effects, axial bending of the pipe in the vicinity of the wavefront, non-planar features of the pressure wavefront, and transient phenomena, may well contribute to the differences between measurement and computation.

Although only direct measurements for the fluid pressure are available, we are able to report estimated data for the fluid velocity as well. 
Since a direct measurement of the fluid velocity is not available, the fluid velocity behind the wavefront is estimated as being equal to the initial velocity of the buffer (see Fig. 3). The initial buffer velocity was estimated by equating the momentum of the projectile prior to impact with that of the buffer after impact. In other experiments using the same experimental setup, high speed video of the projectile-buffer impact has shown that this estimate of the buffer velocity is typically within 10-20\% of its actual velocity.

\begin{table}[h!]
\centering
\caption{Geometrical parameters of pipe 1 (six different cases).
$\dagger=$ aluminum, $\ddagger=$ stainless steel, $\star=$ carbon-epoxy composite. 
} \label{1311101332}
\begin{tabular}{clccccc}
\hline
&\multicolumn{2}{l}{ } 
&\multicolumn{4}{ c}{Speed of primary wave $c_w$ [m/s]} 
\\
%\cline{2-3}
\cline{5-7}
Tube ID    & $R_1$ [m]    & $e_1$ [mm] &  & Measurement &  & Computation\\
%  & [m]    & [mm] & &  [m/s] & & [m/s]\\
\hline
$27^{\dagger}$      & 0.0199   & 1.47 &  & $1245\pm 14.72$  &  & 1207\\
$34^{\dagger}$       & 0.0218      & 0.89  & & $1078\pm 123.50$   &  & 1054\\
$35^{\dagger}$      & 0.0154       & 0.89  & & $1309\pm 36.50$  &  & 1281\\
$38^{\dagger}$     & 0.0154       & 0.89 &  & $1312\pm 46.76$ &  & 1281\\
\hline
$36^{\ddagger}$     & 0.0154      & 0.89  & & $1379\pm 95.92$ &   & 1369\\
\hline
$40^{\star}$    & 0.019     & 1.45 &  & $1157\pm 38.03$ &  & 1100\\
\hline
\end{tabular}
%\caption{Geometrical parameters of pipe 1 (six different cases).
%$\dagger=$ aluminum, $\ddagger=$ stainless steel, $\star=$ carbon-epoxy composite. 
%%For a discussion of the elastic properties of the carbon-epoxy composite see Remark below.
%} \label{1311101332}
\end{table}
\begin{table}[h!]
\centering
\caption{
Comparison of measurements and predictions of maximum hoop strain accompanying the water-hammer wave (pipe 1) and fluid velocity. Here the fluid velocity $V_0^{\dagger}$ was not measured, but instead was estimated at the
post-processing level. 
%Hoop strain $^{\ddagger}$ has been computed from the estimated velocity $V_0^{\dagger}$ by %using Eqs. (\ref{1305091618}) and (\ref{1311100014}) 
%while 
Variables with a superscript $^{\sharp}$ have been computed using the water pressure $P_0$ determined from the experiments and by using Eqs. (\ref{1305091618}) and (\ref{1311100014}). %With some abuse of notation here by $\varepsilon_{1,hoop}$ we denote the maximum hoop %strain accompanying the water-hammer wave.
}
\label{1310091629}
\begin{tabular}{llcccccc}
\hline
&\multicolumn{2}{l}{Experiments} 
&\multicolumn{4}{r}{Computations} 
\\
\cline{2-4}
\cline{6-7}
Shoot ID    & $P_0$  & $\varepsilon_{1,hoop}$   & $V_0^{\dagger}$ & &$\varepsilon_{1,hoop}^{\sharp}$ & $V_0^{\sharp}$ \\
  & [MPa]  &[mstr]  & [m/s] & & [mstr] & [m/s] \\
\hline
100 (tube ID 27)      & 3.10     & -0.61     & 2.86 &  & -0.6581   & 2.57 \\
106 (tube ID 27)       & 6.44    & -1.23     & 5.69  & & -1.3672   & 5.34  \\
%108 (tube ID 27)      & 8.90     & -1.80     & 5.83  & & -1.8895   & 7.38 \\
%\hline
137 (tube ID 34)      & 2.64    & -0.93      & 2.83 &  & -0.9816  & 2.51 \\
151 (tube ID 35)       & 4.98     & -1.37    & 3.97  & &-1.3325  &3.89 \\
200 (tube ID 38)      & 3.05     & -0.55    & 3.14  & & -0.8161 & 2.38 \\
\hline
159 (tube ID 36)      & 7.92     & -0.61     & 5.22 &  & -0.7054   & 5.76 \\
160 (tube ID 36)      & 11.10    & -0.76     & 6.19  & & -0.9886   & 8.07  \\
\hline
398 (tube ID 40)      & 3.12   & -1.56    & - &  & -1.3   & - \\
\hline
\end{tabular}
\end{table}

\begin{figure}[h!]
\centering
%\includegraphics[width=4.7cm]{shep001-eps-converted-to.pdf}%
%\includegraphics[width=4.7cm]{shep002-eps-converted-to.pdf}%
%\includegraphics[width=4.7cm]{shep003-eps-converted-to.pdf}%
%\\
%\includegraphics[width=4.7cm]{shep004-eps-converted-to.pdf}%
%\\
\includegraphics[width=6.5cm, height=4.5cm]{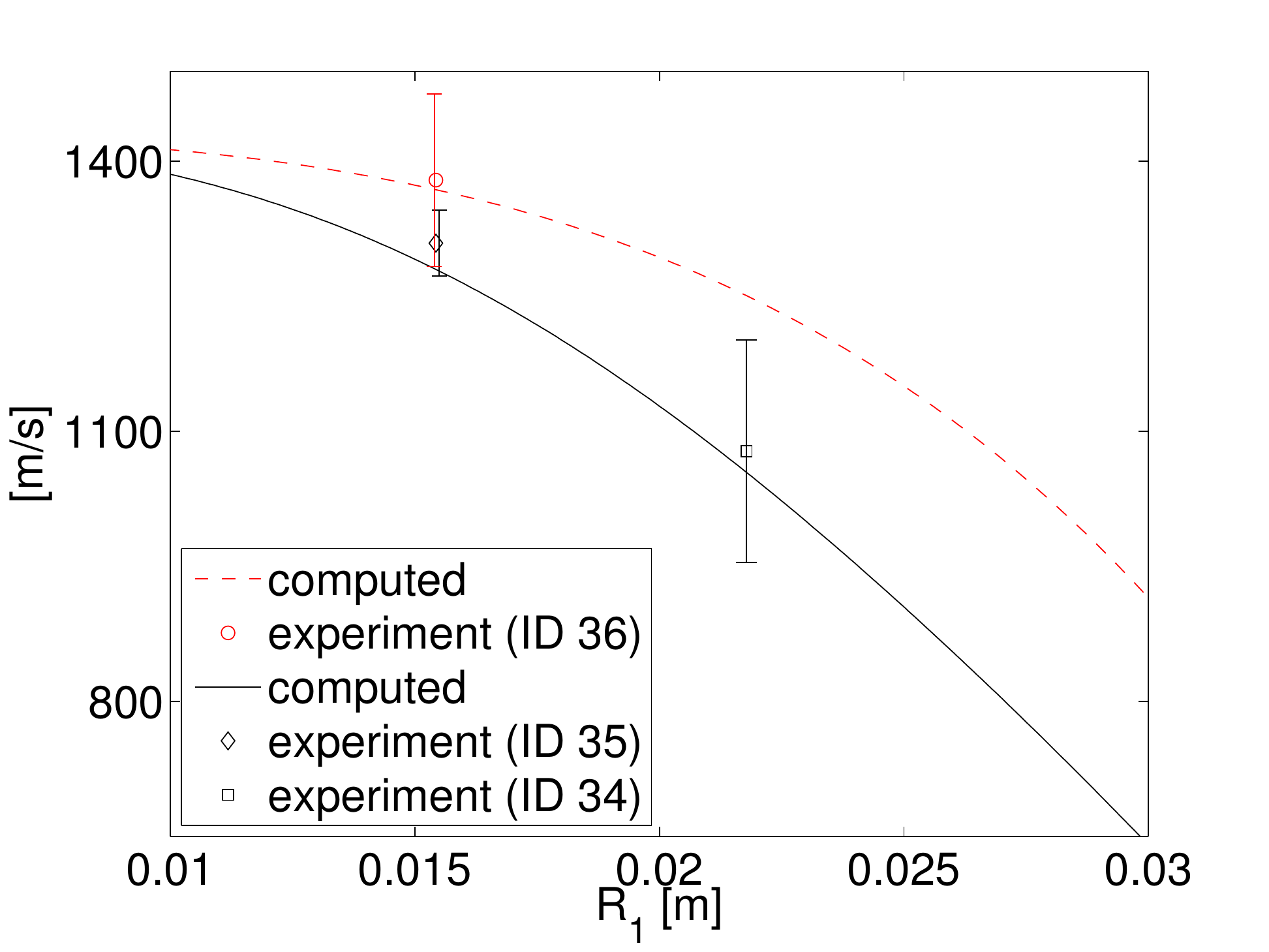}%
\includegraphics[width=6.5cm, height=4.5cm]{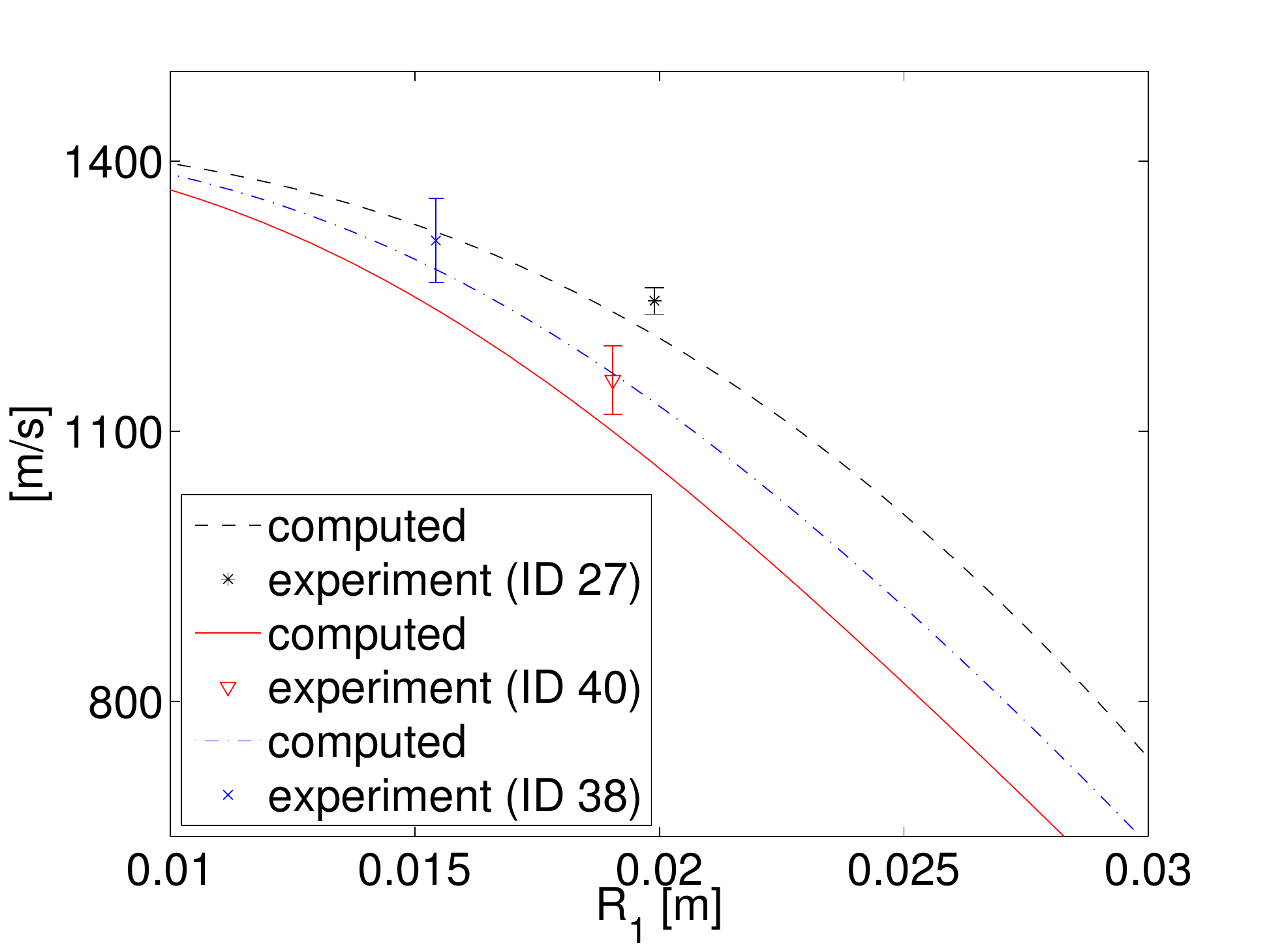}%
\caption{
Comparison of computed primary wave speed $c_w$ and experimental measurements. 
LEFT:
aluminum, tube IDs 34 and 35 (prediction reported as continuous line); stainless steel, tube ID 36 (prediction reported as dashed line).
RIGHT: carbon-epoxy composite, tube ID 40 (prediction reported as a continuous line); aluminum, tube ID 27 (prediction reported as dashed line) and ID 38 (dash-dotted line).
The web version of this article contains the above plot figures in color.
}\label{1406101408}
\end{figure}
 % file pipe 5.m

%\caption{Comparison of computed primary wave speed $c_w$ and experimental measurements. 
%LEFT: aluminum, tube IDs 34, 35 and 38  (prediction reported as continuous line); stainless %steel, tube ID 36 (prediction reported as dashed line).
%RIGHT: carbon-epoxy composite, tube ID 40 (prediction reported as a continuous line); %aluminum, tube ID 27 (prediction reported as dashed line).}

 \paragraph{Remark 1.}

\begin{figure}[h!]
\centering
\includegraphics[width=7cm]{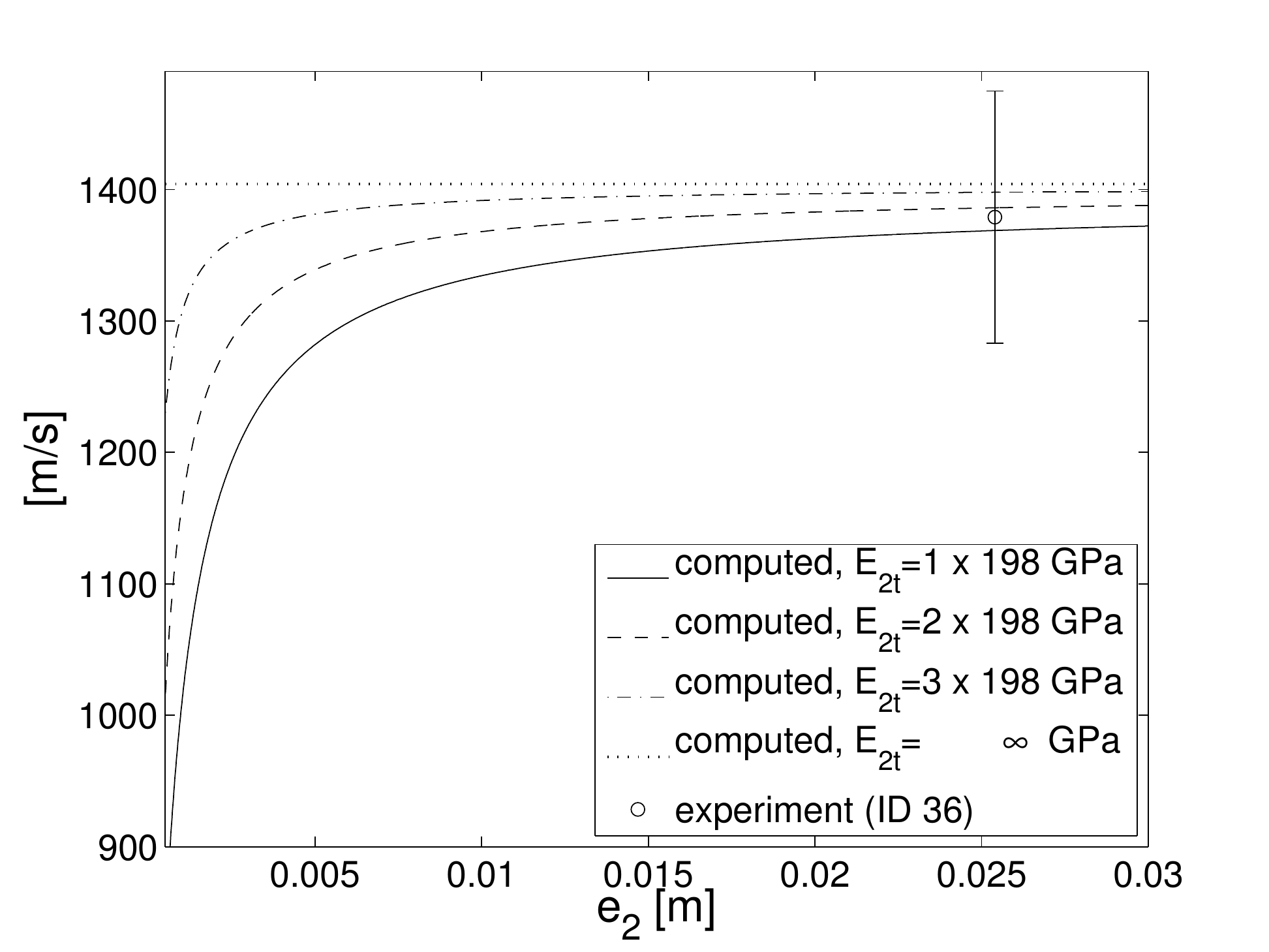}
\includegraphics[width=7cm]{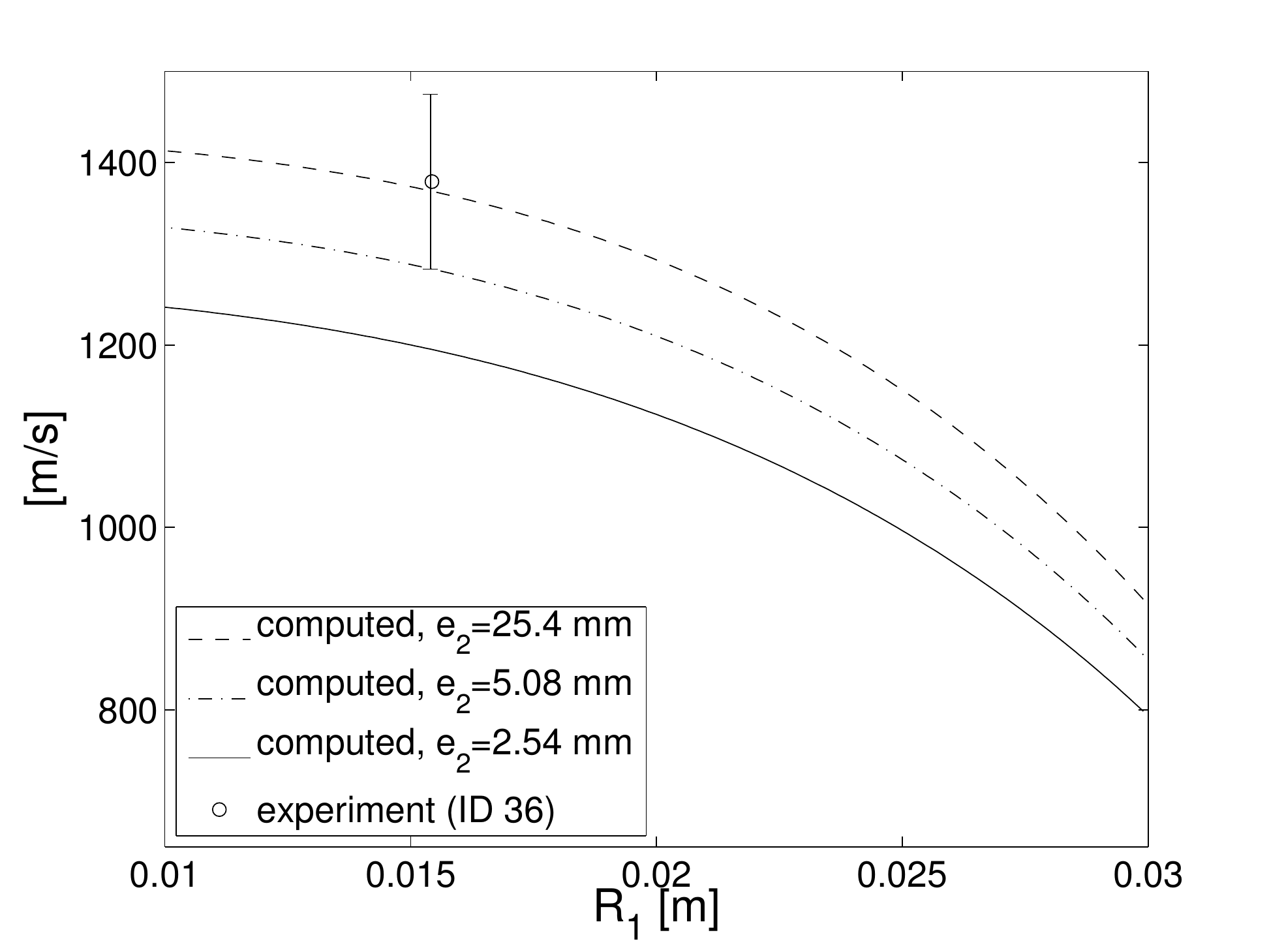}
\caption{
LEFT: Computation of the primary wave speed $c_w$ as a function of the thickness of the outer tube. 
Here various plots correspond to increasing values of the Young's modulus of pipe 2 including the asymptotic case of an ideally rigid outer pipe. The experimental measurement is reported for tube ID 36.
RIGHT:
Computation of the primary wave speed $c_w$ for different values of the thickness of the outer tube. The experimental result available for tube ID 36 is to be compared with the curve obtained for $e_2=25.4$ mm.}\label{1406271625}
\end{figure}
%pipe5s_asy
  
%pipe5thinthin 

While our analysis is valid for pipes of various thicknesses and elastic moduli, experimental results
are available only for a system where thin internal pipes are coupled to a relatively thick
and stiff outer pipe. In this regime, the external pipe is almost rigid and the main contributions to
the fluid-structure interaction are due to the internal pipe. To clarify this point, we analyze the dependence of the primary wave speed on certain parameters of the external pipe. As
an example, we limit our analysis to systems with an internal pipe made of stainless steel with
parameters reported in Table 3. Similar results can be obtained by considering an aluminum
or composite pipe.

%While we refer to an external pipe made of 4140 steel (see Table \ref{1311101234}) here 

%\footnote{ with $R_1=15.4$ mm and $e_1=0.89$ mm, thus corresponding to tub ID 36)}

In Fig. \ref{1406271625}-LEFT we report the computation of the primary wave speed as a function of the thickness of the external pipe and considering the Young's modulus of pipe 2 as a parameter. Therefore the continuous curve corresponds to 4140 steel while the other curves correspond to a material with increasing stiffness including the limit case of an ideally rigid material.
Interestingly, the computed value of the primary wave speed for a 4140 steel pipe with internal radius $R_1=38.1$ mm and thickness of $e_1=25.4$ mm ($c_w=1369$ m/s, Table \ref{1311101332}) is reasonably close to
asymptotic value obtained for a perfectly rigid external pipe ($c_w=1404$ m/s) thus confirming that the external pipe is, with a good approximation, almost rigid.

%
%Interestingly, the computed value of the primary wave speed for a 4140 steel pipe with internal radius $R_1=38.1$ mm and thickness of $e_1=25.4$ mm is reasonably close to
%asymptotic value obtained for a perfectly rigid external pipe. 
%Indeed, comparison with the experimental result  available for tube ID 36 shows that, assuming the outer pipe ideally rigid one obtains an error of 2.5$\%$ while the full model predicts an error of 0.5$\%$ (see Table \ref{1311101332}). This confirms that the external pipe is, with a good approximation, almost rigid.
%

%Fig. \ref{1406271625}-RIGHT contains a graphical representation of the dependence of $c_w$ %on $e_2$ including the case of a thin outer pipe. 

Fig. \ref{1406271625}-RIGHT shows the dependence of $c_w$ on $e_2$ to clarify the effect of a softer outer pipe. 
Again, we consider an internal pipe made of stainless steel and with $e_1=0.89$ mm while the external pipe is in 4140 steel and with fixed $R_2=38.1$ mm. The plot of the wave speed, as a function of $R_1$, generalizes the one 
%for stainless steel 
in Fig. \ref{1406101408}-LEFT as the thickness of the outer pipe is now regarded as a parameter.
 Indeed, for $e_2=25.4 $ mm we obtain the dashed curve plotted in Fig. \ref{1406101408}-LEFT for which a comparison with an experimental result is available. 
For $e_2=2.54$ mm we have $R_2/e_2=15$ which may well be regarded in the regime of thin pipes.
As $e_2$ decreases we obtain decreasing profiles and values of the wave speed and, by Eq. (\ref{1305091618}), of the fluid pressure as well. Indeed, softer pipes undergo large radial deformations 
causing an increase of the annular area and consequently a drop of pressure.
%We refer the interested readers to an upcoming paper concening the detailed description of the %behavior of pipe systems as functions of geometrical parameters.

%infinity wave speed is 1404
% neal dice 1376
%io calcolo 1369

\paragraph{Remark 2.}

The computations of the elastic stiffness for carbon-epoxy composite tubes 
deserves a special comment. When we have more fibers in one direction than in another we must adapt the method used in Paragraph \ref{1312181655}.
Indeed, since tube 40 is a lay-up structure of layers containing fibers in both axial ($\theta=0^o$) and transverse ($\theta=90^o$) direction, its stiffness matrix, denoted by $\overline{\CC}^{ce}_1$, is determined by a volumetric average of the two stiffness matrices obtained by plugging $\theta=0^o$ or $\theta=90^o$ into $\overline{\CC}^{+\theta}$. 
Although the exact proportion of plies with fibers in either direction is unknown, we have been able to estimate the elastic coefficients 
by simply assuming that the known coefficient $\overline{C}_{1,33}^{ce}$
is 
given by a linear combination of $\overline{C}_{33}^{0^o}$ and $\overline{C}_{33}^{90^o}$
\cite{KOLLAR}.
Indeed, by solving the system of equations
\begin{eqnarray}
\overline{C}_{1,11}^{ce}=\xi \,\overline{C}_{11}^{0^o}+(1-\xi) \overline{C}_{11}^{90^o},\qquad
\overline{C}_{1,33}^{ce}=\xi\, \overline{C}_{33}^{0^o}+(1-\xi) \overline{C}_{33}^{90^o},
\end{eqnarray}
we are able to compute the two unknown variables $\overline{C}_{1,11}^{ce}$ and $\xi$. 
The latter, $\xi$, is the relative amount of fibers in the axial direction, a non-dimensional parameter which takes into account both the percentage of plies with fibers \text{in the axial direction} and their thickness. 
For pipes composed of uniaxial layers with all fibers in direction $\theta=0^o$ we have $\xi=1$, while in the dual case of fiber-reinforced pipes with winding angle $\theta=90^o$ we have $\xi=0$.
Notice that the elastic coefficient for CFRP pipes in Paragraph \ref{1312181655} have been computed for $\xi=0.5$ since we have the same amount of plies with winding angle $+\theta$   as those with winding angle $-\theta$ and all the plies have the same thickness.
By plugging $\overline{C}_{33}^{0^o}=\overline{C}_{11}^{90^o}=142$ GPa and $\overline{C}_{11}^{0^o}=\overline{C}_{33}^{90^o}=9$ GPa \cite{private} and $\overline{C}_{1,33}^{ce}=117$ GPa 
\cite{url} yields $\overline{C}_{1,11}^{ce}\approx 34$ GPa and $\xi\approx 0.8$.
%
%\footnote{
%$$
%\xi= \frac{E_{3,t}-E_{1,u}}{E_{3,u}-E_{1,u}}.
%$$
%}
%
Finally, the computation of the density of the composite and of the Poisson's ratio follow as in Eq. (\ref{1312251840}) by assuming
$\rho_{f}=1770$ $\textrm{kg/m}^3$, $\nu_f=0.2$ (carbon fiber) and $\rho_{m}=1208$ $\textrm{kg/m}^3$, $\nu_m=0.39$ (epoxy resin) \cite{HoYou13} with fiber volume fraction $V_f=2/3$ \cite{url} yielding $\nu_{31}=0.26$ and $\rho_{1,t}=1583$ kg/m$^3$.
\\
\quad
\\
Investigation of the pressure waves and mechanical strain profiles becomes particularly transparent when the system is composed of two coaxial carbon-reinforced pipes and when fibers and matrix have the same physical properties in both pipes.
Even though for this case experimental results are not available, in the following we report the analysis for the readers convenience thus generalizing the one-pipe modeling work of \cite{HoYou13}.

\subsection{Fiber-reinforced plastic pipes}\label{1312181514}

The study of the fluid-structure interaction in anisotropic structures, considered for the first time in this paper with the modeling tube ID 40 in Paragraph \ref{1312181656}, is now analyzed in full detail for a system of two coaxial water-filled CFRP pipes composed of pairs of $\pm\theta$ plies of same thickness.  
To keep our analysis simpler, we assume that the physical properties of matrix and fibers (including the winding angle) are identical to one another in both pipes (see Table \ref{1309291458}) yielding $\overline{C}_{1,kl}=\overline{C}_{2,kl}$. As in Paragraph \ref{1312181655} we drop the subscript $i$ in the stiffness and compliance coefficients.
%
%Indeed, with the exception of the pipes radii $R_2>R_1$,
%we assume that all the relevant geometrical properties of fibers and matrix are identical to one %another (see Table \ref{1309291458}). 
%
The general case of pipes with different properties, including different winding angles $\theta_1\neq\theta_2$, can be studied with similar techniques and is left to the brave reader.

The plots of the stiffness and compliance elements $\overline{C}_{kl}$ and $\overline{S}_{kl}$ as a function of the winding angle $\theta$ are contained in \cite{HoYou13}. 
Here it is enough to recall that at $\theta=0^{\circ}$, the direction of the fibers coincides with the axial direction of the pipes and consequently the axial stiffness $\overline{C}_{33}$ has a maximum while the hoop stiffness $\overline{C}_{11}$ has a minimum. As $\theta$ increases, $\overline{C}_{11}$ increases while $\overline{C}_{33}$ decreases due to the fiber reinforcement in the hoop direction. 
%
%The qualitative behavior of $\overline{S}_{11}$ and $\overline{S}_{33}$ is the opposite of those %of $\overline{C}_{11}$ and $\overline{C}_{33}$, respectively. 
%
Conversely, as $\theta$ increases, $\overline{S}_{11}$ decreases while $\overline{S}_{33}$ increases.
Then, the absolute values of the coupling terms $\overline{C}_{13}$ and $\overline{S}_{13}$ have a maximum at $\theta=45^o$ and are minimized at $\theta=0^o$ and $90^o$.
\begin{table}[h!]
\centering
\caption{Geometrical and physical data for coaxial CFRP pipes, \cite{HoYou13}. }\label{1309291458}
\begin{tabular}{llcclcc}
\hline
Carbon fiber &  & &   & Epoxy resin (matrix) &  & 
\\
\hline
Young's modulus  & $E_{f}$  & 238 GPa    &    & Young's modulus      & $E_{m}$  & 2.83 GPa\\
Density  & $\rho_{f}$   &  1770 kg/m$^3$     &  & Density  & $\rho_{m}$ &  1208 kg/m$^3$ \\
Poisson's ratio       & $\nu_{f}$   & 0.2    &     & Poisson's ratio  & $\nu_{m}$ & 0.39  \\
\hline
Composite pipes &  & &   &  &  & 
\\
\hline
Inner radius (pipe 1)  & $R_{1}$  & 19.15 mm   &    & Inner radius (pipe 2)   & $R_2$  & 54 mm\\
Thickness (pipe 1, 2)  & $e_1\hbox{=}e_2$   &  1.66 mm     &  & Fiber volume fraction  & $V_{f}$ &0.7 \\
\hline
\end{tabular}
%\caption{Geometrical and physical data for coaxial CFRP pipes, %\cite{HoYou13}}\label{1309291458}
\end{table}

\begin{figure}[h!]
\centering
\includegraphics[width=6.5cm,height=4.3cm]{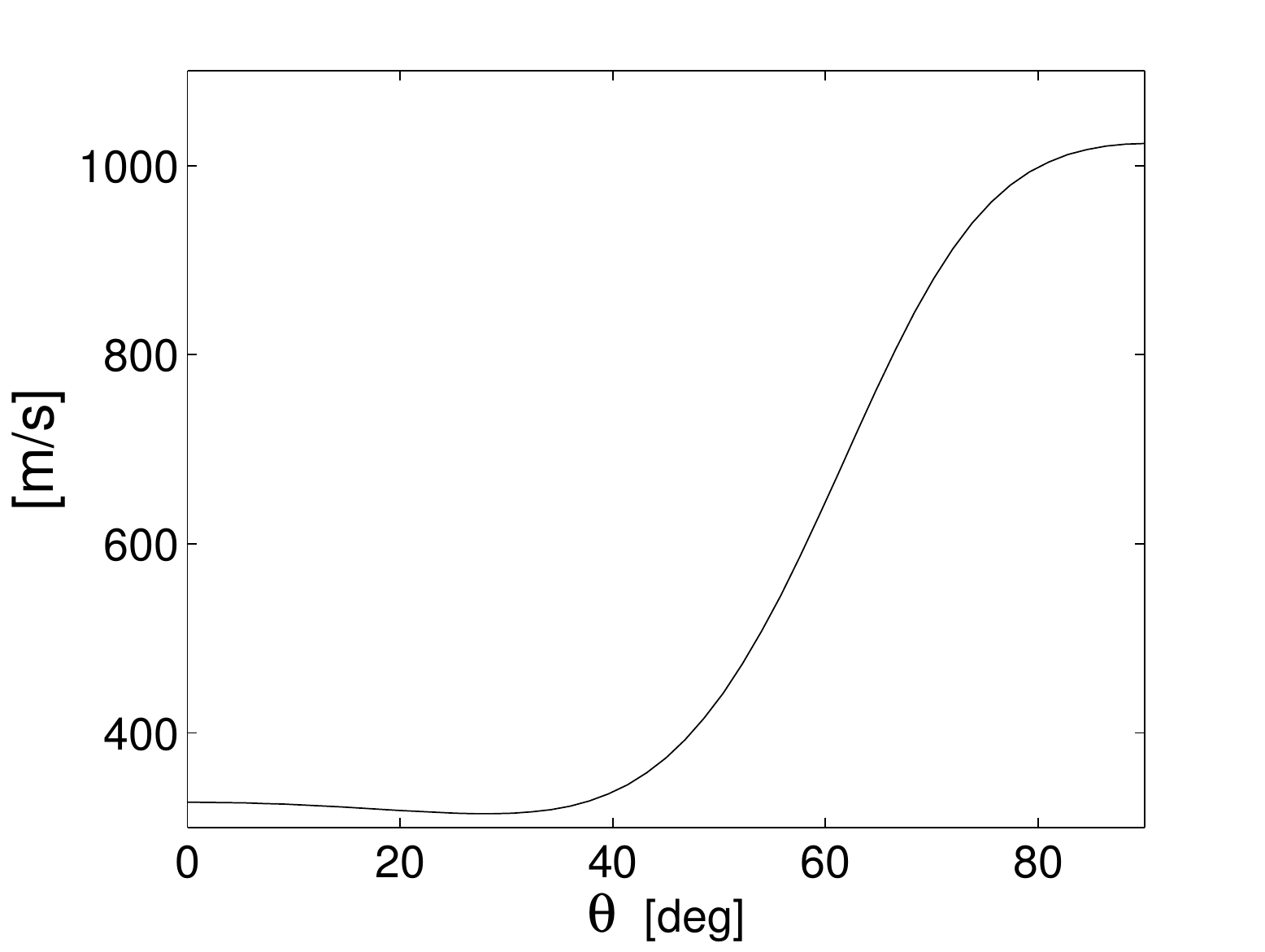}%
\includegraphics[width=6.5cm,height=4.3cm]{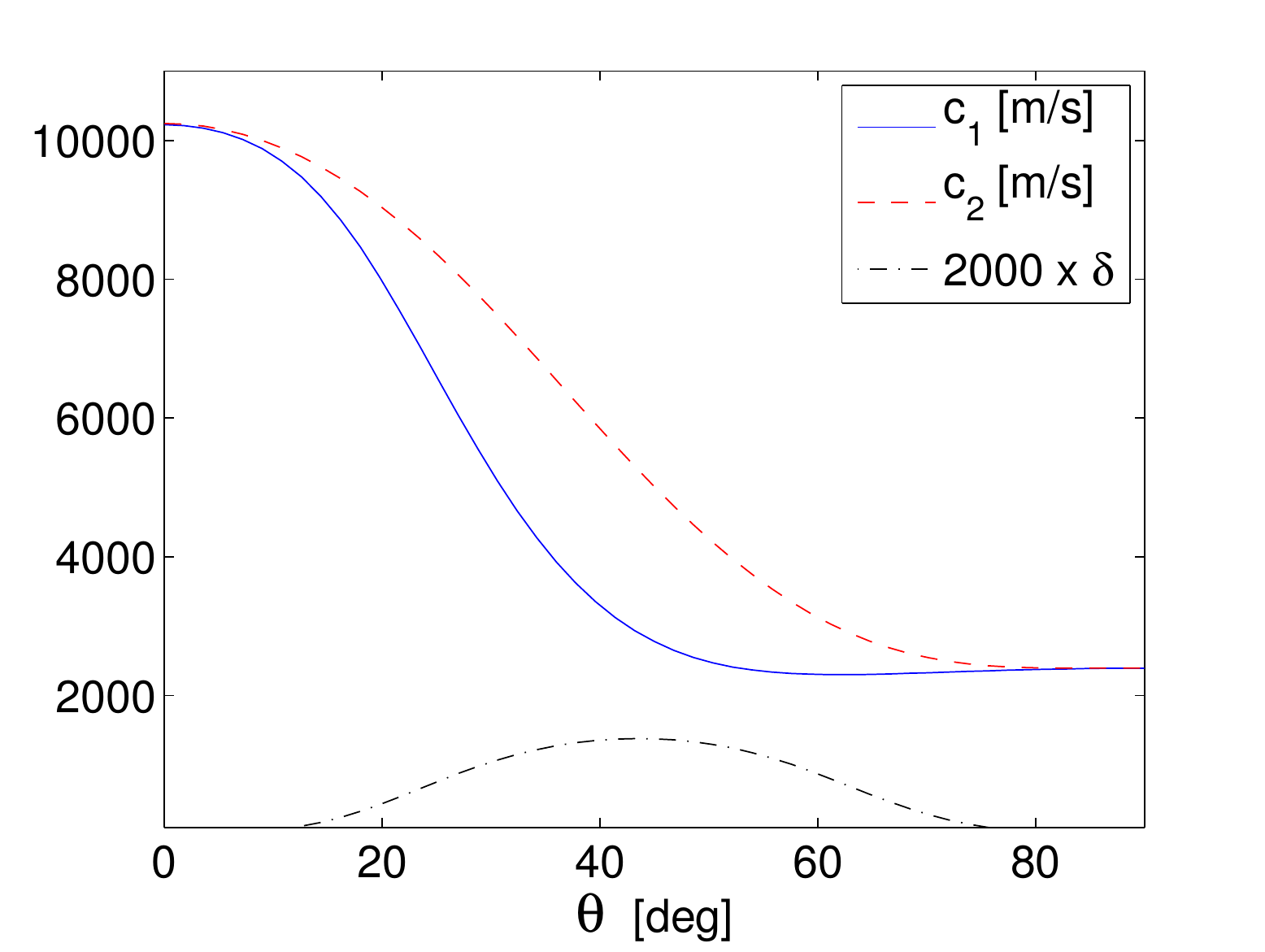}%
\caption{LEFT: plot of the primary wave speed $c_w$, RIGHT: plot of the precursor wave speeds $c_1$ and $c_2$ as a function of the winding angle $\theta$ and of the coefficient $\delta$ (non-dimensional, multiplied by $2000$) as a function of $\theta$. The web version of this article contains the above plot figure in color.}\label{1305101143}
\end{figure}

\begin{figure}[h!]
 \centering
\includegraphics[width=6.5cm, height=4.3cm]{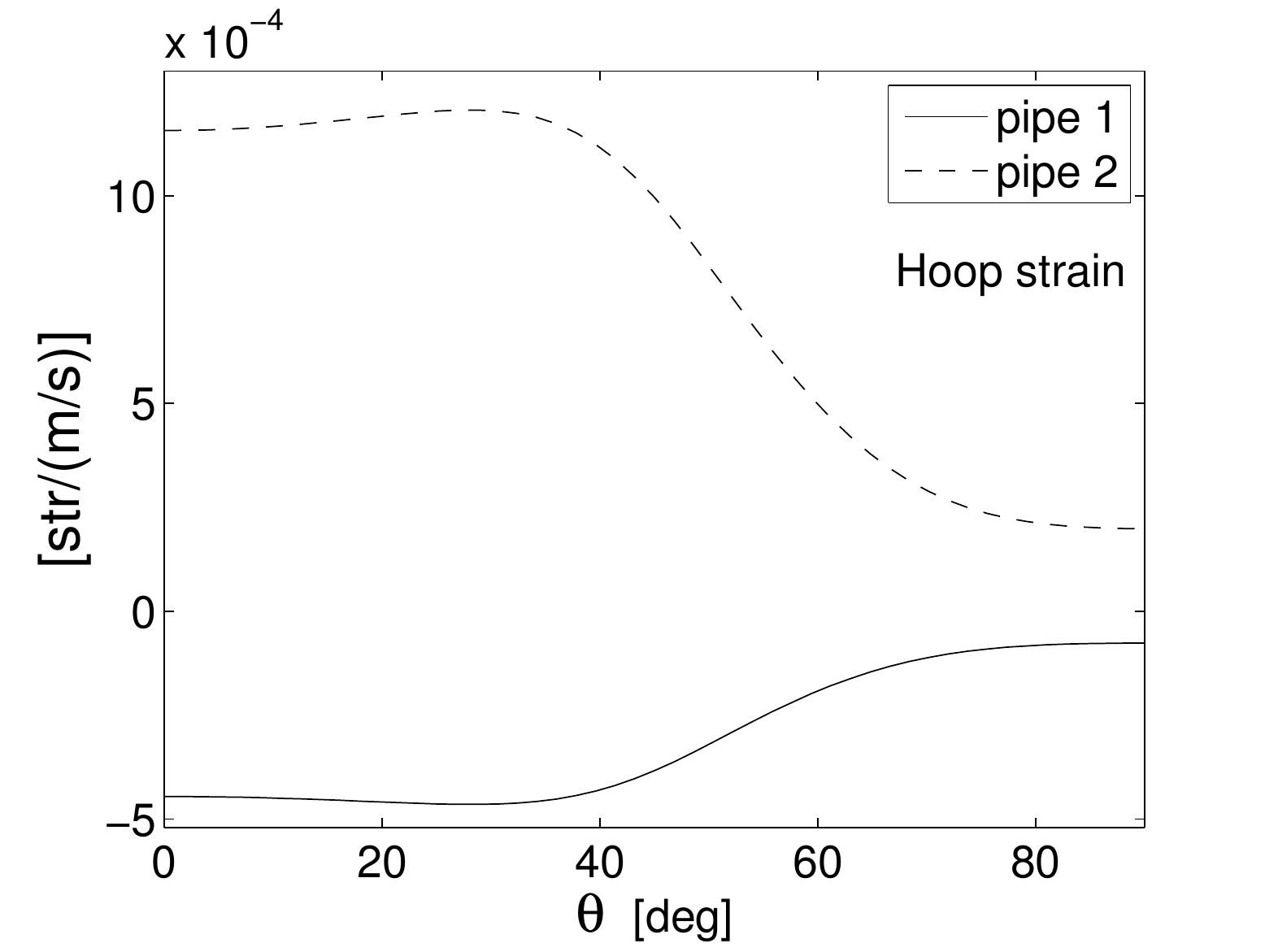}%
\includegraphics[width=6.5cm ,height=4.3cm]{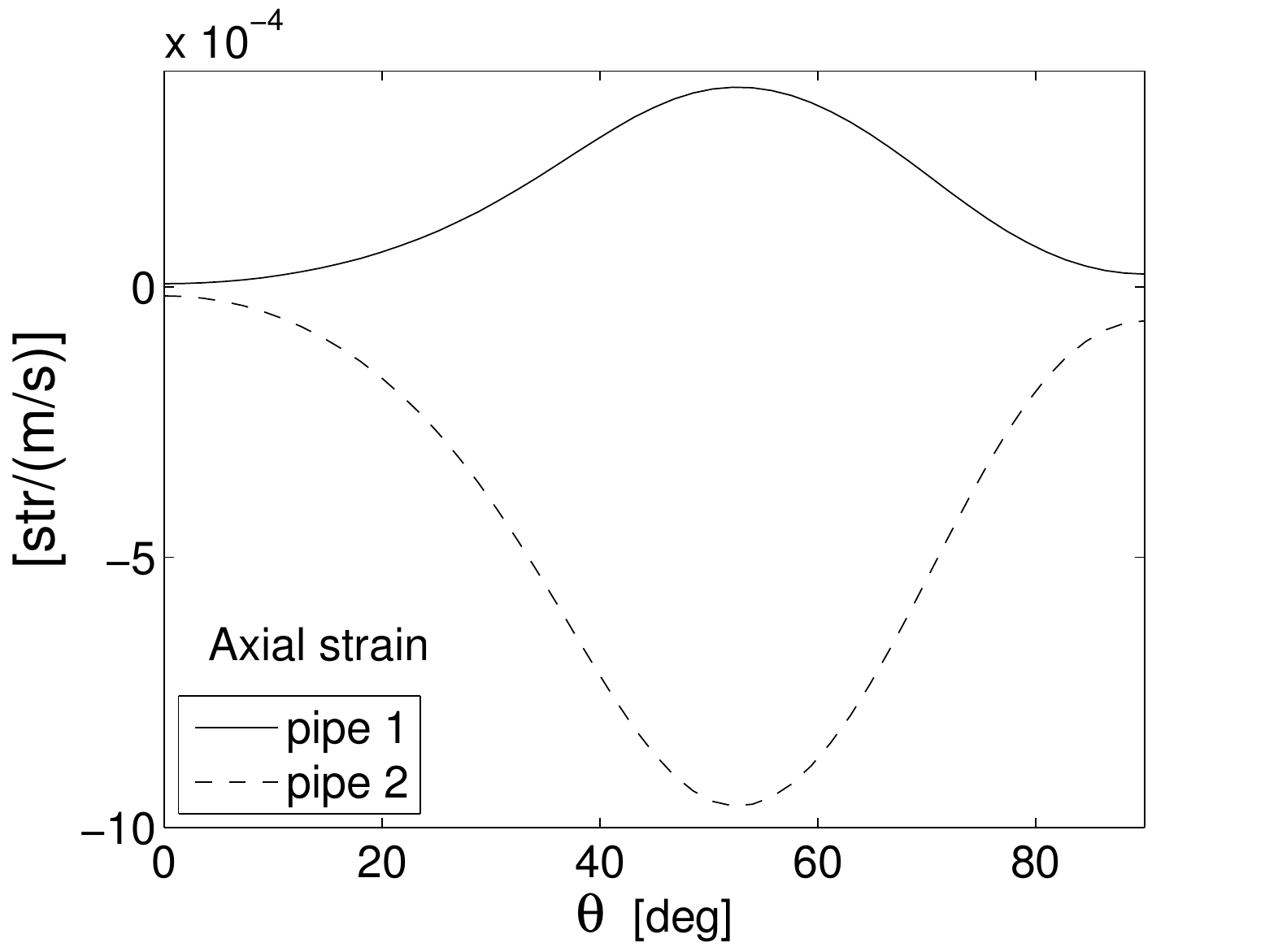}%
\caption{LEFT: plot of the 
maximum hoop strains accompanying the water-hammer wave
normalized by $V_0=1$ m/s as a function of the winding angle $\theta$ (Eqs. (\ref{1311100014})). RIGHT: plot of the 
maximum axial strains accompanying the water-hammer wave
normalized by $V_0=1$ m/s as a function of the winding angle $\theta$ (Eq. (\ref{1309291512})). 
}
\label{1305101157}
\end{figure}

\noindent The calculated primary and precursor waves are shown in Fig. \ref{1305101143} as functions of $\theta$. 
Similar to the one-pipe scenario of \cite{HoYou13}, as $\theta$ increases the result is a reinforcement of the hoop stiffness and an increase in the primary wave speed which corresponds to the breathing mode of the pipes.
Conversely, the speed of the precursor waves (corresponding to a longitudinal mode of the pipes) diminishes due to decreased axial stiffness.
%
%
%(both associated to the incident wave $c_w$)
%As slower waves generate the larger magnitude in both axial and hoop strain, 
%
We remark that the computation of the physical variables is based on the incident wave $c=c_w$ because its magnitude is much larger than those of precursor waves. 
In this case, by Eq. (\ref{1305091618}) we have $P_0/V_0=c_w\rho_w $ and therefore the graph of the fluid pressure is not reported. 
Precursor wave speeds $c_1$ and $c_2$ are very similar when the parameter $\delta$ is very small which happens for either small or large $\theta$. 
This is consistent with Eq. (\ref{1309302314}) where for $\delta=0$ we have $c_1=c_2$. It is possible to prove that
this corresponds to the cases for which the coupling stiffness $\overline{S}_{13}$ as a function of $\theta$ is minimal.
On the other hand, it follows that for intermediate values of $\theta$ wavespeed $c_1$ becomes slower while $\delta$ becomes larger and is maximized at $\theta\approx 43^o$ when also $\overline{S}_{13}$ is large.
%
% 
%A more detailed asymptotic analysis left to a forthcoming paper shows that $\delta$ is strictly %dependent on the coupling term $\overline{S}_{13}$.
%
%
Proof of the dependence of $\delta$ on the coupling term $\overline{S}_{13}$ requires 
a more detailed asymptotic analysis and it is therefore left to a forthcoming paper.
%
% because the precursor waves are essntially the longitudinal waves of the pipes.
%
%
Plots of the hoop and axial strain in pipes 1 and 2 calculated from Eqs. (\ref{1311100014}) and (\ref{1309291512}) are reported in Fig. \ref{1305101157}.
As the impulsive impact by the projectile generates a positive water pressure, pipe 2 undergoes a positive expansion in the radial direction (hoop strain is positive)
while pipe 1 is contracted in the radial direction (hoop strain is negative).
Note that the radial expansion of pipe 2 accompanies the contraction in the axial direction while the radial contraction of pipe 1 accompanies the expansion in the axial direction. This explains why hoop and axial strain have opposite signs.
The hoop strain is essentially determined by the hoop compliance $\overline{S}_{11}$ and therefore the absolute values of the hoop strains in both pipe 1 and 2 are large for small $\theta$ and are decreasing for increasing $\theta$. Then, at a first order of approximation, the axial strain is mainly determined by the coupling term $\overline{S}_{13}$ \cite[Sect. 5]{HoYou13} and therefore the axial strains in both pipe 1 and 2 are 
maximized (in absolute value) for $\theta\approx 52^o$.

\section{Summary and future perspectives}

We have investigated the propagation of stress waves in water-filled pipes in an annular geometry. A six-equation model that describes the fluid-structure interaction has been derived and adapted to both elastically isotropic and anisotropic (fiber-reinforced) pipes. The natural frequencies of the system (eigenvalues) and the amplitude of the pressure and velocity of the fluid, along with the mechanical strains and stresses in the pipes (eigenvectors) have been computed and compared with experimental data of water-hammer tests. 
%\textbf{Our results show good agreement even when certain geometrical assumptions on the %pipe wall thickness are not satisfied.} 
It is observed that the projectile impact causes a positive expansion of the external pipe in the radial direction and a contraction in the axial direction (Poisson's effect). Vice versa, the internal pipe is contracted in the radial direction and expanded axially.
In the last section of the paper, which is a benchmark for future experimental investigation, we have analyzed in full detail the propagation of waves in CFRP pipes with a special emphasis on the influence of the winding angle on the 
%combined fluid-structure response of the system. 
wave speeds and the axial and hoop strains in the pipes.
Most interestingly, we found that the speed of the primary wave (breathing mode) increases with the increasing winding angle due to increasing hoop stiffness in both pipes. This is in agreement with the one-pipe model and analysis of \cite{HoYou13}. Conversely, the profile of the speed of the precursor waves (longitudinal modes) is large when the winding angle is small (and therefore the axial stiffness is large) and it decreases with increasing winding angle.
Additionally, we have observed that the two precursor waves travel at almost the same speed when the fiber winding angle is equal to either $0^o$ or $90^o$ while a separation of the velocities is observed for $\theta$ in between these values, a phenomenon which is the object of further analysis.

\subsection*{Acknowledgments}
The authors are indebted to Prof. K. Bhattacharya and J. Shepherd for many useful discussions and their advice.
P.C. acknowledges support from the Department of Energy 
National Nuclear Security Administration under Award Number 
DE-FC52-08NA28613.
This work was written when P.C. was a postdoctoral student at
the California Institute of Technology.

%\subsubsection*{Stiffness and compliance matrix for fiber-reinforced pipes}

\end{document}